\documentclass[12pt]{article}
\usepackage{amssymb}
\usepackage{latexsym}
\usepackage{amsmath}
\usepackage{graphicx}

\usepackage{esdiff}
\usepackage{mathtools}

\def\vp{\varphi}
\def\ve{\varepsilon}
\def\al{\alpha}
\def\nb{\nabla}
\def\th{\theta}
\def\e{\eta}
\def\l{\lambda}
\def\k{\kappa}
\def\L{\Lambda}
\title{\LARGE{Topological black hole in the theory with nonminimal derivative coupling with power-law Maxwell field and its thermodynamics}}
\author{M. M. Stetsko\footnote{E-mail: mstetsko@gmail.com}\
\\
  {\small Department for Theoretical Physics, Ivan Franko National University of Lviv,}\\
{\small 12 Drahomanov Str., Lviv, UA-79005, Ukraine
         }}

\begin{document}
\maketitle

{\abstract{We obtain topological black hole solutions in scalar-tensor gravity with nonminimal derivative coupling between scalar and tensor components of gravity and power-law Maxwell field minimally coupled to gravity. The obtained solutions can be treated as a generalization of previously derived charged solutions with standard Maxwell action \cite{Feng_PRD16}. We examine the behaviour of obtained metric functions for some asymptotic values of distance and coupling. To obtain information about singularities of the metrics we calculate Kretschmann scalar. We also examine the behaviour of gauge potential and show that it is necessary to impose some constraints on parameter of nonlinearity in order to obtain reasonable behaviour of the filed. The next part of our work is devoted to the examination of black hole's thermodynamics. Namely we obtain black hole's temperature and investigate it in general as well as in some particular cases. To introduce entropy we use well known Wald procedure which can be applied to quite general diffeomorphism-invariant theories. We also extend thermodynamic phase space by introducing thermodynamic pressure related to cosmological constant and as a result we derive generalized first law and Smarr relation. The extended thermodynamic variables also allow us to construct Gibbs free energy and its examination gives important information about thermodynamic stability and  phase transitions. We also calculate heat capacity of the black holes which demonstrates variety of behaviour for different values of allowed parameters.}}
\section{Introduction}
General Relativity is very renowned theory that gives explanations for numerous facts and is in agreement with lots of observations  starting from planetary scales and up to the cosmological ones \cite{Will_LRR2014,Berti_CQG2015}. One of the biggest prediction of General Relativity that was made in early years of this theory, namely gravitational waves, finally have been discovered experimentally \cite{Abbott_PRL16,Abbott_PRL16_2,Abbott_PRL17}  after the century of great ideas but unsuccessful attempts to detect them. Despite the great success of General Relativity there are still several open issues which are extremely important in order to have more complete picture of our Universe. Among those puzzles which remain unsolved yet we distinguish the problems about the origin of cosmological singularities, cosmological constant problem and closely related to it dark energy and dark matter issues and also the question about the presence of terms of higher order in curvature than the standard Einstein-Hilbert term. All the mentioned questions bring the idea of modification of General Relativity \cite{Clifton_PhysRept2012,Heisenberg_2018}, but there are a lot of possible ways of this generalization and some of them have received great interest and have been applied to numerous problems of Gravity and Cosmology, the others have not become so popular and known. Systematic description of open issues and the present day challenges of General Relativity and the approaches which face these challenges and give reasonable solutions for them is given in quite recent review \cite{Heisenberg_2018}. 

Amidst these generalizations of standard General Relativity that have got broad interest are the so called scalar-tensor theories, which can be classified by the type of coupling between gravitational tensor and scalar degrees of freedom. We would also like to distinguish Horndeski gravity which appears to be the most general scalar-tensor theory where the equations of motion for tensor as well as scalar components are of the second order \cite{Horndeski_IJTP74}.  Being almost unknown for more than three decades it has been studied intensively since the early years of the last decade. This interest to some extent was motivated by the examination of the so called Galileon theories which are a sort of the scalar-tensor theories with a specific Galilean invariance regarding the scalar field \cite{Deffayet_PRD09}.  It was shown that Galileon theory and Horndeski gravity are equivalent \cite{Kobayashi_PTEP11}. Galileon action was also obtained with help of the standard Kaluza-Klein procedure of compactification of higher dimensional Lovelock theory \cite{Acoleyen_PRD11,Charmousis_LNP15}. It should be pointed out that here dilaton fields which appear due to the use of the Kaluza-Klein procedure takes the role of the Galileons. We also note that some particular cases of general Horndeski theory, namely the theory with nonminimal derivative coupling, were derived in low energy String Theory \cite{Metsaev_NPB87,Meissner_PLB97,Cartier_PRD01}. Galileon theory also appears in the framework of massive gravity \cite{deRham_PRD11}. It should be pointed out that more general multiscalar versions of Horndeski theory were formulated \cite{Deffayet_PRD10,Padilla_JHEP13,Charmousis_JHEP14,Ohashi_JHEP15} and theories beyond Horndeski were considered \cite{Zumala_PRD14,Gleyzes_PRL15,Gleyzes_JCAP15,Langlois_JCAP16,Chrisostomi_JCAP16,BenAchour_PRD16,BenAchour_JHEP16}.

Horndeski gravity and its particular cases have been applied to numerous problems of cosmology and black holes' physics. Namely, exact cosmological solutions within the nonominimal derivative coupling theory were examined \cite{Sushkov_PRD09}. Cosmological scenarios including quintessence and phantom fields were investigated and transition between different types of de Sitter solutions was shown \cite{Saridakis_PRD10}. Models of dark energy/dark matter in the framework of nonminimally coupled gravity and scalar field were considered and their dynamics was examined \cite{Granda_JCAP10, Gao_JCAP10,Sadjadi_PRD11}. Influence of power-law scalar potential on cosmological dynamics was investigated \cite{Skugoreva_PRD13}. New inflationary mechanisms which were supposed to be more realistic were examined \cite{Sushkov_PRD12}. Nonminimally coupled scalar-tensor theory can successfully generate slow rolling inflation without violation of unitarity bounds \cite{Germani_PRL10,Germani_PRL11}. Reheating process during rapid oscillations of the inflation and curvaton scenario in nonminimal derivative coupling models were studied \cite{Sadjadi_JCAP13,Dalianis_JCAP17,Feng_PLB14,Feng_PRD14,Qiu_EPJC17}. Reconstruction of Horndeski gravity from effective field theory of dark energy was also investigated \cite{Kennedy_PRD17}. Constraints on Horndeski theory and its predictions in the light of modern gravitational observation were also discussed \cite{Hou_EPJC17,Crisostomi_PRD18,Langlois_PRD18,Heisenberg_2018}. 

Black holes and other compact astrophysical objects are also very interesting area of investigation. Taking into account quite recent observation of  gravitational waves and the fact that the sources of the waves are the compact objects they become of extreme importance for the verification of predictions of different generalizations of standard of General Relativity, including Horndeski theory, its generalizations or particular cases. Due to complicated structure of general Horndeski theory the special interest was devoted to the examination of black holes in some particular cases of Horndeski theory. Namely, a static black hole solution in the model with nonminimal derivative coupling but without cosmological constant was obtained \cite{Rinaldi_PRD12}. Four and five-dimensional solutions in the the theory with cosmological constant was found and some aspects of its thermodynamics were investigated \cite{Minamitsuji_PRD14}. More general multidimensional and topological solutions were obtained \cite{Anabalon_PRD14,Cisterna_PRD14,Kobayashi_PTEP14}. Stealth Schwarzschild and de Sitter self-tuned solutions were found \cite{Babichev_JHEP14}. Slowly rotating neutron stars and black holes in the nonminimal derivative coupling sector of Horndeski gravity were examined in \cite{Cisterna_PRD16,Maselli_PRD15,Stetsko}. Three dimensional solutions were considered in \cite{Bravo-Gaete_PRD14,Giribet_PRD15,Clement_arxiv}. Some aspects of black hole thermodynamics were studied \cite{Feng_JHEP15,Feng_PRD16,Stetsko}.  Black holes in more general Lovelock Galileon theory were investigated \cite{Charmousis_PRD15}. Stability of black holes in Horndeski gravity was paid attention to in the papers \cite{Cisterna_PRD15,Takahashi_PRD17,Tretyakova_CQG17,Ganguly_CQG18,Babichev_PRL18}. Causal structure of black holes in shift-symmetric Horndeski theories was examined \cite{Benkel_arxiv}. Some aspects of neutron and boson stars in Horndeski gravity were considered \cite{Cisterna_PRD16,Cisterna_PRD15,Brihaye_PRD16,Verbin_PRD18}. The existence of black hole's hair in Horndeski gravity was studied \cite{Sotiriou_PRL14}. Hairy black holes in the theory with nonminimal coupling and Gauss-Bonnet term were investigated \cite{Antoniou_PRL18}. 

As it has been noted above charged topological black holes in Einstein-Horndeski gravity with standard (linear) Maxwell action term were considered and their thermodynamics was studied in the paper \cite{Feng_PRD16}.  In our work we examine topological black holes in the theory with nonminimal derivative coupling between tensor and scalar components and nonlinear Maxwell field with Lagrangian of power-law type. The obtained in our work black holes' solutions can be considered as generalization of previously mentioned solutions \cite{Feng_PRD16} and when the factor (power) of nonlinearity $p=1$  those previously obtained solutions \cite{Feng_PRD16} are recovered. It is known that standard Maxwell Lagrangian in a space-time of dimension other than four is not conformally invariant and its generalization to power-law nonlinear expression allows to recover the conformal invariance \cite{Hassaine_PRD07}. We point out here that Maxwell term with the power-law nonlinearity in Lagrangian was considered in different works related to black holes \cite{Hassaine_PRD07,Maeda_PRD09,Gurtug_PRD12,Kord_PRD15,Kord_PRD15_2}. Having obtained the black hole solutions we investigate their thermodynamics, namely we calculate temperature, entropy, Gibbs free energy and heat capacity and study their behaviour. We note that some keypoints of the present work were discussed in the work \cite{Stetsko_2}.

This work is organized as follows. In the second section we derive topological black hole solutions and investigate the behaviour of metric functions for different distances and regimes of coupling, we also examine Kretschmann scalar in order to make some conclusions about singular points of the obtained solution, we also obtain gauge potential and derive some constraints on parameter of nonlinearity in order to have this potential reasonable. The third section which is divided on several subsections is devoted to the study of thermodynamic properties of the obtained black  holes. Namely using standard geometric relation we calculate black hole's temperature, then with help of Wald procedure we obtain a relation for entropy and it allows us to construct the first law of black hole's thermodynamics. Using extended thermodynamics approach we calculate Gibbs free energy and derive the generalized first law and Smarr relation, we also investigate heat capacity of the black holes. Finally, the forth section contains some conclusions.

\section{Equations of motion for the system with nonlinear Maxwell field}
We consider an action for a system which can be described by a standard Einstein-Hilbert term with cosmological constant and some scalar field with nonminimal derivative coupling and we also suppose that the additional Maxwell field we take into account is represented by a nonlinear term. So, the action integral can be written in the form:
\begin{equation}\label{action}
S=\int d^{n+1}x\sqrt{-g}\left( R-2\Lambda-\frac{1}{2}\left(\alpha g^{\mu\nu}-\eta G^{\mu\nu}\right)\partial_{\mu}\vp\partial_{\nu}\vp +\left(-F_{\mu\nu}F^{\mu\nu}\right)^p\right)
\end{equation}
where $g_{\mu\nu}$  denotes the metric tensor, $g=det{g_{\mu\nu}}$ is the determinant of the metric, $G_{\mu\nu}$ and $R$ are the Einstein tensor and Ricci scalar correspondingly, $\L$ is the cosmological constant and $\vp$ is the scalar field, nonminimally coupled to gravity and finally $F_{\mu\nu}$ denotes the Maxwell field tensor and $p$ is the factor (power) of nonlinearity and when $p=1$ the standard Maxwell action term is recovered.

Having used the principle of the least action we derive equations of motion which can be written in the form:
\begin{equation}\label{eom}
G_{\mu\nu}+\Lambda g_{\mu\nu}=\frac{1}{2}(\alpha T^{(1)}_{\mu\nu}+\eta T^{(2)}_{\mu\nu})+T^{(3)}_{\mu\nu}
\end{equation}
and here we use the following notations
\begin{equation}\label{scal_min}
T^{(1)}_{\mu\nu}=\nb_{\mu}\vp\nb_{\nu}\vp-\frac{1}{2}g_{\mu\nu}\nb^{\lambda}\vp\nb_{\lambda}\vp,
\end{equation}
\begin{eqnarray}\label{scal_nm}
\nonumber T^{(2)}_{\mu\nu}=\frac{1}{2}\nb_{\mu}\vp\nb_{\nu}\vp R-2\nb^{\lambda}\vp\nb_{\nu}\vp R_{\lambda\mu}+\frac{1}{2}\nb^{\lambda}\vp\nb_{\lambda}\vp G_{\mu\nu}-g_{\mu\nu}\left(-\frac{1}{2}\nb_{\lambda}\nb_{\kappa}\vp\nb^{\lambda}\nb^{\kappa}\vp\right.\\\left.+\frac{1}{2}(\nb^2\vp)^2-R_{\lambda\kappa}\nb^{\lambda}\vp\nb^{\kappa}\vp\right)
-\nb_{\mu}\nb^{\lambda}\vp\nb_{\nu}\nb_{\lambda}\vp+
\nb_{\mu}\nb_{\nu}\vp\nb^2\vp-R_{\lambda\mu\kappa\nu}\nb^{\lambda}\vp\nb^{\kappa}\vp
\end{eqnarray}
\begin{equation}\label{max_tr_nlin}
T^{(3)}_{\mu\nu}=\frac{g_{\mu\nu}}{2}\left(-F_{\l\k}F^{\l\k}\right)^p+2p\left(-F_{\l\k}F^{\l\k}\right)^{p-1}F_{\mu\rho}{F_{\nu}}^{\rho}
\end{equation}
It should be noted that the term $T^{(1)}_{\mu\nu}$ is the standard energy-momentum tensor for a scalar field minimally coupled to gravity whereas the collection of terms denoted by $T^{(2)}_{\mu\nu}$ is caused by the nonminimal coupling to gravity. We also remark that the terms denoted by $T^{(3)}_{\mu\nu}$ comprise the energy-momentum tensor for nonlinear Maxwell field and it is easy to check that when $p=1$ one arrives at the standard Maxwell field energy-momentum tensor. Variation of the action (\ref{action}) with respect to the scalar field $\vp$ gives rise to a corresponding equation of motion of the form:
\begin{equation}\label{scal_f_eq}
(\alpha g_{\mu\nu}-\eta G_{\mu\nu})\nb^{\mu}\nb^{\nu}\vp=0.
\end{equation}
And for the Maxwell field one obtains:
\begin{equation}\label{Maxwell_eq}
\nb_{\mu}\left((-F_{\l\k}F^{\l\k})^{p-1}F^{\mu\nu})\right)=0.
\end{equation}
One can verify that when $p=1$ the standard Maxwell equations can be recovered.

We are to obtain so called topological solutions and we assume that the metric we find takes the form:
\begin{equation}\label{metric}
ds^2=-U(r)dt^2+W(r)dr^2+r^2d\Omega^{2(\ve)}_{(n-1)},
\end{equation}
where $d\Omega^{2(\ve)}_{n-1}$ is the line element of a $n-1$-dimensional hypersurface of a constant curvature and it takes the form. 
\begin{eqnarray}
d\Omega^{2(\ve)}_{(n-1)}=
\begin{cases}
d\th^2+\sin^2{\th}d\Omega^2_{(n-2)}, \quad \ve=1,\\
d\th^2+{\th}^2 d\Omega^2_{(n-2)},\quad \ve=0,\\
d\th^2+\sinh^2{\th}d\Omega^2_{(n-2)},\quad \ve=-1,
\end{cases}
\end{eqnarray}
and here $d\Omega^2_{(n-2)}$ is the line element of a $n-2$--dimensional sphere. The given expression for $d\Omega^{2(\ve)}_{(n-1)}$ represents surfaces of positive, null or negative curvature for given values of parameter $\ve$. 

Having supposed that the electromagnetic field has a scalar component only ($A=A_0(r)dt$) and taking into consideration the chosen form of the metric (\ref{metric}) one can solve the Maxwell equations (\ref{Maxwell_eq}) and obtain electromagnetic field tensor in the form:
\begin{equation}\label{EM_field}
F_{rt}=-\frac{q}{r^{(n-1)/(2p-1)}}\sqrt{UW}
\end{equation}
and here $q$ is an integration constant related to the charge of the black hole.
Now we consider the equation for the scalar field (\ref{scal_f_eq}) and having integrated it for a once we derive:
\begin{equation}
\sqrt{\frac{U}{W}}r^{n-1}\left[\al-\e\frac{(n-1)}{2rW}\left(\frac{U\rq{}}{U}-\frac{(n-2)}{r}(\ve W-1)\right)\right]\vp'=C
\end{equation}
We assume that the constant $C$ is equal to zero and this condition is equivalent to the following relation:
\begin{equation}
\al g_{rr}-\e G_{rr}=0
\end{equation}
Having used the metric (\ref{metric}) we can write the equations (\ref{eom}) in the following form:
\begin{eqnarray}\label{G_0}
\nonumber \frac{(n-1)}{2rW}\left(\frac{W\rq{}}{W}+\frac{(n-2)}{r}(\ve W-1)\right)\left(1+\frac{3}{4}\e\frac{(\vp\rq{})^2}{W}\right)-\L=\frac{\al}{4W}(\vp\rq{})^2+\\\frac{\e}{2}\left(\frac{(n-1)}{rW^2}{\vp''}\vp\rq{}+\frac{(n-1)(n-2)}{r^2W^2}(\vp\rq{})^2\left(\ve W-\frac{1}{2}\right)\right)+\frac{2p-1}{2}\left(\frac{2q^2}{r^{2(n-1)/(2p-1)}}\right)^p
\end{eqnarray}
\begin{eqnarray}\label{G_1}
\nonumber \frac{(n-1)}{2rW}\left(\frac{U\rq{}}{U}-\frac{(n-2)}{r}(\ve W-1)\right)\left(1+\frac{3}{4}\e\frac{(\vp\rq{})^2}{W}\right)+\L=\\\frac{\al}{4W}(\vp\rq{})^2-\frac{\e}{2}\ve\frac{(n-1)(n-2)}{2r^2W}(\vp\rq{})^2-\frac{2p-1}{2}\left(\frac{2q^2}{r^{2(n-1)/(2p-1)}}\right)^p
\end{eqnarray}

\begin{eqnarray}\label{G_2}
\nonumber\left[\frac{1}{2UW}\left(U\rq{}\rq{}-\frac{(U\rq{})^2}{2U}-\frac{U\rq{}W\rq{}}{2W}\right)+\frac{n-2}{2rW}\left(\frac{U\rq{}}{U}-\frac{W\rq{}}{W}\right)-\frac{(n-2)(n-3)}{2r^2W}(\ve W-1)\right]\times \\\nonumber\left(1+\frac{\e}{4}\frac{(\vp\rq{})^2}{W}\right)+\L=-\frac{\al}{4W}(\vp\rq{})^2-\frac{\e}{2W^2}\vp\rq{}\rq{}\vp\rq{}\left(\frac{U\rq{}}{2U}+\frac{n-1}{r}\right)+\\\frac{\e}{2}\frac{(\vp\rq{})^2}{W}\left(\frac{U\rq{}W\rq{}}{4UW^2}+\frac{(n-2)W\rq{}}{2rW^2}-\ve\frac{(n-2)(n-3)}{2r^2}\right)+\frac{1}{2}\left(\frac{2q^2}{r^{2(n-1)/(2p-1)}}\right)^p
\end{eqnarray}
Having combined  the equations (\ref{G_0}) and (\ref{G_1}) we arrive at the relations:
\begin{equation}\label{vp_der_sq}
(\vp\rq{})^2=-\frac{4r^2W}{\e(2\al r^2+\ve\e(n-1)(n-2))}\left(\al+\L\e+\frac{2^{p-1}(2p-1)\e q^{2p}}{r^{2p(n-1)/(2p-1)}}\right)
\end{equation}
\begin{equation}\label{UW_prod}
UW=\frac{\left((\al-\L\e)r^2+\ve\e(n-1)(n-2)-2^{p-1}\e(2p-1)q^{2p}r^{2(1-p(n-1)/(2p-1))}\right)^2}{(2\al r^2+\ve\e(n-1)(n-2))^2}.
\end{equation}
It is worth being emphasized that the square of derivative of the scalar field should be positive definite in the domain outside black hole's horizon (since we are to obtain black hole's solution). To provide the positive definiteness of $(\vp')^2$ a condition on the parameters of coupling $\al$, $\e$ and cosmological constant $\L$ should be imposed, for example for positive $\al$ and $\e$ one should take the cosmological constant $\L$ to be a negative one since the metric function $W(r)$ is supposed to be positive definite in the outer domain (if the parameter $\e$ is negative it can be shown that the negative value of the cosmological constant also should be taken negative).  In the inner domain (inside the black hole) the right hand side of the relation (\ref{vp_der_sq}) might change its sign and it means that the scalar field might turn to be a phantom-like field, but the behaviour of the field $\vp$ inside the black hole does not affect on its behaviour in the outer domain. We would also like to point out that the right hand side of the relation (\ref{UW_prod}) is always nonnegative and it means that the sign of the metric functions $U(r)$ and $W(r)$ is always the same (when one of them changes its sign, so does the other function), as it has to be for a black hole. It should be noted that when $p=1$ the given above relations (\ref{vp_der_sq}) and (\ref{UW_prod}) reduce to the corresponding relations, obtained for linear Maxwell field \cite{Feng_PRD16} .
 The metric function $U(r)$ can be represented in the form:
 \begin{eqnarray}\label{U_int}
\nonumber U(r)=\ve-\frac{\mu}{r^{n-2}}-\frac{2\L}{n(n-1)}r^{2}-2^p\frac{(2p-1)^2q^{2p}}{(n-1)(2p-n)}r^{2\left(1-\frac{p(n-1)}{2p-1}\right)}+\\\nonumber\frac{(\al+\L\e)^2}{2\al\e(n-1)r^{n-2}}\int\frac{r^{n+1}}{r^2+d^2}dr+2^{p-1}\frac{(2p-1)(\al+\L\e)q^{2p}}{\al(n-1)r^{n-2}}\times\\
\int\frac{r^{n+1-\frac{2p(n-1)}{2p-1}}}{r^2+d^2}dr+2^{2p-3}\frac{(2p-1)^2\e q^{4p}}{\al(n-1)r^{n-2}}\int\frac{r^{n+1-\frac{4p(n-1)}{2p-1}}}{r^2+d^2}dr
 \end{eqnarray}
 and here $d^2=\ve\e(n-1)(n-2)/2\al$.
 The integrals in the written above formula gives rise to different expressions in case of the exponents of the $r$ in the numerators of the integrals are integer or not. The form of the final expression for the metric function $U(r)$ also depends on the sign of the introduced parameter $d^2$ and the relation between integration variable $r$ and this parameter. The first of the mentioned integrals always has positive integer exponent, so when $d^2>0$ it can be represented in the form:
\begin{equation}\label{i_1}
\int \frac{r^{n+1}}{r^2+d^2}dr=(-1)^{(n+1)/2}d^n\arctan\left({\frac{r}{d}}\right)+\sum^{(n-1)/2}_{j=0}(-1)^jd^{2j}\frac{r^{n-2j)}}{n-2j}
\end{equation}
for odd values of $n$ and
\begin{equation}\label{i_2}
\int \frac{r^{n+1}}{r^2+d^2}dr=(-1)^{n/2}\frac{d^n}{2}\ln\left(\frac{r^2}{d^2}+1\right)+\sum^{n/2-1}_{j=0}(-1)^jd^{2j}\frac{r^{n-2j)}}{n-2j}
\end{equation}
for the case of even $n$. One can conclude that the first integral in the formula (\ref{U_int}) can be expressed in terms of elementary functions consisting of power terms and a transcendental function which depends on the fact whether the dimension of space $n$ is odd or even. 

The second and the third integrals in the right hand side of the relation (\ref{U_int}) might have nonpositive integer or noninteger exponents in the numerators. It should be pointed out here that the second integral might have positive exponent when $p>(n+1)/4$, but the exponent of the third integral always takes negative values.  In the case when $\frac{2p(n-1)}{2p-1}$ is integer both exponents are integer and the integrals can be represented in the form:
\begin{equation}\label{int_ni_even}
\int \frac{r^{[N_i]}}{r^2+d^2}dr=\frac{(-1)^{-\frac{[N_i]}{2}}}{d^{1-[N_i]}}\arctan{\left(\frac{r}{d}\right)}+\sum^{-\frac{[N_i]}{2}-1}_{j=0}(-1)^j\frac{r^{[N_i]+2j+1}}{([N_i]+2j+1)d^{2(j+1)}}
\end{equation} 
when $[N_i]$ is even and
\begin{equation}\label{int_ni_odd}
\int \frac{r^{[N_i]}}{r^2+d^2}dr=\frac{(-1)^{-\frac{([N_i]-1)}{2}}}{2d^{1-[N_i]}}\ln{\left(1+\frac{d^2}{r^2}\right)}+\sum^{-\frac{([N_i]+3)}{2}}_{j=0}(-1)^j\frac{r^{[N_i]+2j+1}}{([N_i]+2j+1)d^{2(j+1)}}
\end{equation} 
for odd $[N_i]$, and here we denoted $N_1=n+1-\frac{2p(n-1)}{2p-1}$ and $N_2=n+1-\frac{4p(n-1)}{2p-1}$ and the brackets $[N_i]$ mean that the parameter $N_i$ ($i=1,2$) takes an integer value. In general nor $N_1$ neither $N_2$ are integers, so the corresponding integrals can be expressed only through the special functions. In particular, for large $r$ (large $r$ means that $r>b$) we can write:
\begin{equation}\label{i_3}
\int \frac{r^{N_i}}{r^2+d^2}dr=r^{N_i-1}\sum^{+\infty}_{j=0}\frac{1}{N_i-1-2j}\left(-\frac{d^2}{r^2}\right)^j=\frac{r^{N_i-1}}{(N_i-1)}{_{2}F_{1}}\left(1,\frac{1-N_i}{2};\frac{3-N_i}{2};-\frac{d^2}{r^2}\right)
\end{equation}
The series written above is convergent, when $\frac{d^2}{r^2}<1$. In contrast for small distances, namely, when ${r}<d$  another representation of the hypergeometric function can be used, so the latter integral can be rewritten in the form:
\begin{equation}\label{i_4}
\int \frac{r^{N_i}}{r^2+d^2}dr=\frac{r^{N_i+1}}{(1+N_i)d^2}{_{2}F_{1}}\left(1,\frac{1+N_i}{2};\frac{3+N_i}{2}; -\frac{r^2}{d^2}\right)
\end{equation}
It is clear that when $N_i$ is integer the given hypergeometric functions can be rewritten through the elementary logarithmic or inverse tangent functions.
Using the derived forms for integrals we can write the general relation for the metric function. Namely, for odd $n$ and noninteger $N_1$ and $N_2$ we obtain:
 \begin{eqnarray}\label{funct_odd}
\nonumber U(r)=\ve-\frac{\mu}{r^{n-2}}-\frac{2\L}{n(n-1)}r^2+\frac{(\al+\L\e)^2}{2\al\e(n-1)}\left[(-1)^{(n+1)/2}\frac{d^n}{r^{n-2}}\arctan\left({\frac{r}{d}}\right)+\sum^{(n-1)/2}_{j=0}(-1)^jd^{2j}\frac{r^{2(1-j)}}{n-2j}\right]\\\nonumber-2^p\frac{(2p-1)^2q^{2p}}{(2p-n)(n-1)}r^{\frac{2(3p-pn-1)}{2p-1}}+2^{p-1}\frac{(2p-1)^2(\al+\L\e)q^{2p}}{\al(n-1)(2p-n)}r^{2\left(1-\frac{p(n-1)}{2p-1}\right)}\times\\\nonumber{_{2}F_{1}}\left(1,\frac{p(n-1)}{2p-1}-\frac{n}{2};\frac{p(n-1)}{2p-1}-\frac{n}{2}+1;-\frac{d^2}{r^2} \right)+2^{2p-3}\frac{(2p-1)^3\e q^{4p}}{\al(n-1)(4p-2pn-n)}r^{2\left(1-\frac{2p(n-1)}{2p-1}\right)}\times\\{_{2}F_{1}}\left(1,\frac{2p(n-1)}{2p-1}-\frac{n}{2};\frac{2p(n-1)}{2p-1}-\frac{n}{2}+1;-\frac{d^2}{r^2} \right). 
 \end{eqnarray}
For even $n$ we arrive at the following expression:
\begin{eqnarray}\label{funct_even}
\nonumber U(r)=\ve-\frac{\mu}{r^{n-2}}-\frac{2\L}{n(n-1)}r^2+\frac{(\al+\L\e)^2}{2\al\e(n-1)}\left[(-1)^{n/2}\frac{d^n}{2r^{n-2}}\ln\left(\frac{r^2}{d^2}+1\right)+\sum^{n/2-1}_{j=0}(-1)^jd^{2j}\frac{r^{2(1-j)}}{n-2j}\right]\\\nonumber-2^p\frac{(2p-1)^2q^{2p}}{(2p-n)(n-1)}r^{\frac{2(3p-pn-1)}{2p-1}}+2^{p-1}\frac{(2p-1)^2(\al+\L\e)q^{2p}}{\al(n-1)(2p-n)}r^{2\left(1-\frac{p(n-1)}{2p-1}\right)}\times\\\nonumber{_{2}F_{1}}\left(1,\frac{p(n-1)}{2p-1}-\frac{n}{2};\frac{p(n-1)}{2p-1}-\frac{n}{2}+1;-\frac{d^2}{r^2} \right)+2^{2p-3}\frac{(2p-1)^3\e q^{4p}}{\al(n-1)(4p-2pn-n)}r^{2\left(1-\frac{2p(n-1)}{2p-1}\right)}\times\\{_{2}F_{1}}\left(1,\frac{2p(n-1)}{2p-1}-\frac{n}{2};\frac{2p(n-1)}{2p-1}-\frac{n}{2}+1;-\frac{d^2}{r^2} \right).
 \end{eqnarray}
 It should be noted that in the written above expressions for the metric function $U(r)$ the relation (\ref{i_3}) is used and it allows to describe the behaviour of the solution for large $r$ (when $r>b$), whereas for small $r$ ($r<b$), the relation (\ref{i_4}) should be used. 

For linear electromagnetic field ($p=1$) the relations for the metric function $U(r)$ obtained in the paper \cite{Feng_PRD16} are restored, namely for the odd $n$ it can be represented in the form:
\begin{eqnarray}\label{odd_lin}
\nonumber U(r)=\ve-\frac{\mu}{r^{n-2}}-\frac{2\L}{n(n-1)}r^2+\frac{(\al+\L\e)^2}{2\al\e(n-1)}\left[(-1)^{\frac{n+1}{2}}\frac{d^n}{r^{n-2}}\arctan\left({\frac{r}{d}}\right)+\sum^{(n-1)/2}_{j=0}(-1)^jd^{2j}\frac{r^{2(1-j)}}{n-2j}\right]\\\nonumber+\frac{2q^2}{(n-1)(n-2)}r^{2(2-n)}+\frac{(\al+\L\e)q^2}{\al(n-1)r^{n-2}}\left[\frac{(-1)^{\frac{n-3}{2}}}{d^{n-2}}\arctan\left({\frac{r}{d}}\right)+\sum^{(n-5)/2}_{j=0}(-1)^j\frac{r^{4-n+2j}}{(4-n+2j)d^{2(j+1)}}\right]\\+\frac{\e q^4}{2\al(n-1)r^{n-2}}\left[\frac{(-1)^{\frac{3n-5}{2}}}{d^{3n-4}}\arctan\left({\frac{r}{d}}\right)+\sum^{(3n-7)/2}_{j=0}(-1)^j\frac{r^{6-3n+2j}}{(6-3n+2j)d^{2(j+1)}}\right]
\end{eqnarray}
 and for even $n$ we arrive at:
\begin{eqnarray}\label{even_lin}
\nonumber U(r)=\ve-\frac{\mu}{r^{n-2}}-\frac{2\L}{n(n-1)}r^2+\frac{(\al+\L\e)^2}{2\al\e(n-1)}\left[(-1)^{\frac{n}{2}}\frac{d^n}{2r^{n-2}}\ln\left({\frac{r^2}{d^2}}+1\right)+\sum^{n/2-1}_{j=0}(-1)^jd^{2j}\frac{r^{2(1-j)}}{n-2j}\right]\\\nonumber+\frac{2q^2}{(n-1)(n-2)}r^{2(2-n)}+\frac{(\al+\L\e)q^2}{\al(n-1)r^{n-2}}\left[\frac{(-1)^{\frac{n-2}{2}}}{2d^{n-2}}\ln\left(1+\frac{d^2}{r^2}\right)+\sum^{n/2-3}_{j=0}(-1)^j\frac{r^{4-n+2j}}{(4-n+2j)d^{2(j+1)}}\right]\\+\frac{\e q^4}{2\al(n-1)r^{n-2}}\left[\frac{(-1)^{\frac{3n}{2}}}{2d^{3n-4}}\ln\left(1+\frac{d^2}{r^2}\right)+\sum^{3n/2-4}_{j=0}(-1)^j\frac{r^{6-3n+2j}}{(6-3n+2j)d^{2(j+1)}}\right].
\end{eqnarray}
It should be pointed out here that for the linear case $p=1$ as well as for the cases when $\frac{2p(n-1)}{2p-1}$ is integer then the relations (\ref{int_ni_even}) and (\ref{int_ni_odd}) are used and the resulting form of the metric function $U(r)$ contains only power-law and elementary transcendental functions.
 
When $\ve=0$ (flat horizon surface)  the metric function can be written in the following way:
\begin{eqnarray}\label{flat_metr}
\nonumber U(r)=-\frac{\mu}{r^{n-2}}+\frac{(\al-\L\e)^2}{2\al\e n(n-1)}r^2-2^{p-1}\frac{(\al-\L\e)(2p-1)^2q^{2p}}{\al(n-1)(2p-n)}r^{2\left(1-\frac{p(n-1)}{2p-1}\right)}\\+2^{2p-3}\frac{(2p-1)^3\e q^{4p}}{\al(n-1)(4p-n-2pn)}r^{2\left(1-\frac{2p(n-1)}{2p-1}\right)}
\end{eqnarray}

We investigate the behaviour of the metrics (\ref{funct_odd}) and (\ref{funct_even}) for some asymptotic values of $r$ and $\eta$. Firstly, we examine asymptotic  behaviour of metric function $U(r)$ when $r\rightarrow 0$. To perform this task we decompose  all the functions comprising the mentioned function $U(r)$ into a series for small distances. As a result, for small $r\rightarrow 0$ the function can be written in the form:
\begin{eqnarray}\label{metr_f_small_r}
\nonumber U(r)=\ve-\frac{\mu}{r^{n-2}}-\frac{2\L}{n(n-1)}r^2-2^p\frac{(2p-1)^2q^{2p}}{(2p-n)(n-1)}r^{2\left(1-\frac{p(n-1)}{2p-1}\right)}+\frac{(\al+\L\e)^2}{\ve\e^2(n-1)^2(n^2-4)}r^4\\\nonumber+\frac{2^{p}(2p-1)^2(\al+\L\e)q^{2p}}{\ve\e(n-1)^2(n-2)(6p-n-2)}r^{2\left(2-\frac{p(n-1)}{2p-1}\right)}+\frac{2^{2(p-1)}(2p-1)^3}{\ve(n-1)^2(n-2)}\times\\\frac{q^{4p}}{2p(4-n)-n-2}r^{4\left(1-\frac{p(n-1)}{2p-1}\right)}+{\cal O}_1(r^6)+{\cal O}_2\left(r^{2\left(3-\frac{p(n-1)}{2p-1}\right)}\right)+{\cal O}_3\left(r^{2\left(3-\frac{2p(n-1)}{2p-1}\right)}\right)
\end{eqnarray}
and here ${\cal O}_i(r^{\gamma_i})$ denotes the terms that appear as the following terms in the decomposition of corresponding functions in $U(r)$. It should be noted that the first four terms in the expansion (\ref{metr_f_small_r}) completely recover the standard solution for nonlinearly charged black hole with a cosmological constant, but the relation (\ref{UW_prod}) does not give rise to the condition $U(r)W(r)\simeq 1$ which takes place in the standard Einsteinian theory.

Having used asymptotic expansions of the functions in the relations (\ref{funct_odd}) and (\ref{funct_even}) for when $r\rightarrow\infty$ we can write the expressions for the metric function $U(r)$ when the distance $r$ tends to infinity, namely for odd $n$ we arrive at:
\begin{eqnarray}\label{large_r_odd}
\nonumber U(r)=\ve\left(1-\frac{(\al+\L\e)^2}{4\al^2}\right)+\frac{(\al-\L\e)^2}{2n(n-1)\al\e}r^2-\frac{\mu}{r^{n-2}}+\frac{(\al+\L\e)^2}{2(n-1)\al\e}\left((-1)^{\frac{n+1}{2}}\frac{\pi d^{n}}{2 r^{n-2}}+\right.\\\nonumber\left.\sum^{(n-1)/2}_{j=2}(-1)^jd^{2j}\frac{r^{2(1-j)}}{n-2j}\right)-2^{p-1}\frac{(2p-1)^2(\al-\L\e)q^{2p}}{\al(n-1)(2p-n)}r^{2\left(1-\frac{p(n-1)}{2p-1}\right)}+\\\frac{2^{2p-3}(2p-1)^3\e q^{4p}}{\al(n-1)(4p-n-2pn)}r^{2\left(1-\frac{2p(n-1)}{2p-1}\right)}+{\cal O}_1\left(\frac{1}{r^{n-1}}\right)+{\cal O}_2\left(\frac{1}{r^{\frac{2p(n-1)}{2p-1}}}\right)+{\cal O}_3\left(\frac{1}{r^{\frac{4p(n-1)}{2p-1}}}\right),
\end{eqnarray}
and for even $n$ we obtain:
\begin{eqnarray}\label{large_r_even}
\nonumber U(r)=\ve\left(1-\frac{(\al+\L\e)^2}{4\al^2}\right)+\frac{(\al-\L\e)^2}{2n(n-1)\al\e}r^2-\frac{\mu}{r^{n-2}}+\frac{(\al+\L\e)^2}{2(n-1)\al\e}\left((-1)^{\frac{n}{2}}\frac{d^{n}}{r^{n-2}}\ln\left(\frac{r}{d}\right)+\right.\\\nonumber\left.\sum^{n/2-1}_{j=2}(-1)^jd^{2j}\frac{r^{2(1-j)}}{n-2j}\right)-2^{p-1}\frac{(2p-1)^2(\al-\L\e)q^{2p}}{\al(n-1)(2p-n)}r^{2\left(1-\frac{p(n-1)}{2p-1}\right)}+\\\frac{2^{2p-3}(2p-1)^3\e q^{4p}}{\al(n-1)(4p-n-2pn)}r^{2\left(1-\frac{2p(n-1)}{2p-1}\right)}+{\cal O}_1\left(\frac{1}{r^{n}}\right)+{\cal O}_2\left(\frac{1}{r^{\frac{2p(n-1)}{2p-1}}}\right)+{\cal O}_3\left(\frac{1}{r^{\frac{4p(n-1)}{2p-1}}}\right),
\end{eqnarray}
and here ${\cal O}_i$ denotes the subleading terms in the expansions. It should be noted that the behaviour of the leading term of the metric function $U(r)$ at infinity does not depend on choice of the parameter $\ve$ and is completely the same for three types of geometry and this fact has complete analogy with topological black holes in the framework of the standard General Relativity. So, we can conclude that the behaviour of the metric functions at the infinity is deSitterian or anti-deSitterian which depends on the sign of $\e$.  The other moment should be paid attention to is the fact that the leading terms related to the charge (those which contain the parameter $q$) is the same for all three types of geometry. It should also be stressed that the asymptotic behaviour of the metric function when $r\rightarrow 0$ and $r\rightarrow\infty$ is very important for the calculation of Kretschmann scalar which gives information about true and coordinate singularities of the metric.

We also examine the regime of large nonminimal derivative  coupling (large $\e$), this situation means that the term with minimal coupling (the term with factor $\al$) is considerably smaller than the term which represents nonminimal coupling (the term with factor $\e$). For this situation the metric function $U(r)$ can be represented as follows:
\begin{eqnarray}\label{large_eta}
\nonumber U(r)=\ve-\frac{\mu}{r^{n-2}}-\frac{2\L}{n(n-1)}r^2-\frac{2^p(2p-1)^2q^{2p}}{(2p-n)(n-1)}r^{2\left(1-\frac{p(n-1)}{2p-1}\right)}+\frac{\L^2}{\ve(n-1)^2(n^2-4)}r^4+\\\frac{2^p(2p-1)^2\L q^{2p}}{\ve(n-1)^2(n-2)(6p-n-2)}r^{2\left(2-\frac{p(n-1)}{2p-1}\right)}+\frac{2^{2(p-1)}(2p-1)^3 q^{4p}}{\ve(n-1)^2(n-2)(8p-2pn-n-2)}r^{4\left(1-\frac{p(n-1)}{2p-1}\right)}+{\cal O}\left(\frac{1}{\e}\right)
\end{eqnarray}
The latter relation can be compared with the relation (\ref{flat_metr}) obtained for the case of flat horizon geometry and as it is easy to verify that the flat geometry metric contains the terms with the dependence $\sim\e$ whereas for the case of nonflat geometry ($\ve=\pm 1$) such kind of terms does not appear.
The product of the metric functions (\ref{UW_prod}) for the regime of large $\e$ takes the form:
\begin{equation}
UW=\left(1-\frac{\L r^2}{\ve(n-1)(n-2)}-\frac{2^{p-1}(2p-1)q^{2p}r^{2\left(1-\frac{p(n-1)}{2p-1}\right)}}{\ve(n-1)(n-2)}\right)^2+{\cal O}\left(\frac{1}{\e}\right).
\end{equation}
The written above relation is again valid when $\ve=\pm 1$ and for the flat-type geometry ($\ve=0$) the product of metric functions in the regime of large $\e$ is as follows:
\begin{equation}
UW=\left(\L^2+2^p(2p-1)\L q^{2p}r^{-\frac{2p(n-1)}{2p-1}}+2^{2(p-1)}(2p-1)^2q^{4p}r^{-\frac{4p(n-1)}{2p-1}}\right)\frac{\e^2}{4\al^2}+{\cal O}(\e)
\end{equation}
which goes in agreement with the fact that the function $U(r)$ behaves as $\sim\e$ in this regime.
\begin{figure}
\centerline{\includegraphics[scale=0.33,clip]{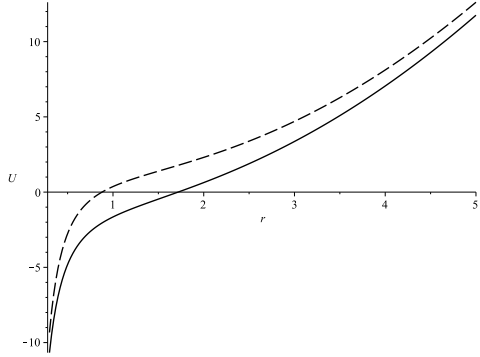}}
\caption{Metric functions $U(r)$ for linear ($p=1$, solid  curve) and conformal ($p=\frac{n+1}{4}$, dashed curve) types of gauge field. All the other parameters are equal for both cases, namely $n=4$, $\al=0.2$, $\e=0.4$, $\ve=1$, $\L=-2$, $q=0.2$ and $\mu=1$.}\label{metr_f_graph}
\end{figure}

The figure Fig.[\ref{metr_f_graph}] shows the behaviour of the metric function $U(r)$ for linear ($p=1$) and conformal cases ($p=\frac{n+1}{4}$). In general the behaviour of these functions is very similar. Both of them go up to infinity with increasing of $r$ and the difference between them becomes negligibly small because of the fact that at the large distances the dominant terms in both functions are  of AdS-type which have the same fixed parameters. It should be pointed out that this conclusion is valid for any allowed value of $p$. For small distances ($r\rightarrow 0$) the metric function for linear case goes faster to minus infinity than in the conformal case. The quite general conclusion we can make here is the fact that with increasing of the parameter of nonlinearity $p$ the behaviour of the metric function for $r\rightarrow 0$ becomes weaker than for smaller value of $p$. 

In order to find out where the metric has physical singularities we should analyze the behaviour of Kretschmann scalar. For our types of metrics it can be written in the form:
\begin{equation}\label{Kr_scalar}
R_{\mu\nu\k\l}R^{\mu\nu\k\l}=\frac{1}{UW}\left(\frac{d}{dr}\left[\frac{U'}{\sqrt{UW}}\right]\right)^2+\frac{(n-1)}{r^2W^2}\left(\frac{(U')^2}{U^2}+\frac{(W')^2}{W^2}\right)+\frac{2(n-1)(n-2)}{r^4W^2}(\ve W-1)^2.
\end{equation}
The represented form of the Kretschmann scalar is suitable for all three types of geometry and certainly the evident form of this scalar would depend on it but for simplicity we have not substituted the final forms of the functions $U$ and $W$ into the written expression of the scalar (\ref{Kr_scalar}).

Having written the Kretschmann scalar (\ref{Kr_scalar}) one can analyze its behaviour for different values of $r$. First of all it is necessary to investigate its behaviour at the horizon points, namely at the point $r_{+}$ where  $U(r_{+})=0$ and the function $W(r)$ is singular. Taking into account the relation (\ref{UW_prod}) it is easy to persuade oneself that the horizon points are the points of coordinate singularity, where the Kretschmann scalar has finite values that is usually takes place for black hole's solution. When $r\rightarrow 0$ we arrive at:
\begin{eqnarray}\label{Kr_scal_origin}
R_{\mu\nu\k\l}R^{\mu\nu\k\l}\sim\frac{16(n-2)^2(3p-pn-1)^2}{(8p-2pn-n-2)^2}\left(\frac{(4p-2pn-1)^2}{(2p-1)^2}+\frac{n(n-1)}{n-2}\right)\frac{1}{r^4}
\end{eqnarray}
and as a consequence a physical singularity appears at the origin. It is worth being noted that for all of dimensions $n$ the character of singularity is identical what is in contrast for example with the case of chargeless black hole, where the character of this singularity depends on the dimension of space \cite{Stetsko} or with the case of a charged black hole in the standard General Relativity where this dependence also takes place. This peculiarity can be explained by the fact that in the case we consider now the product of the metric functions $UW$ is also singular what does not happen nor in the chargeless case \cite{Stetsko} neither in the standard GR, where we have $UW=1$. When $r\rightarrow\infty$ we find:
\begin{equation}\label{Kr_sc_infty}
\nonumber R_{\mu\nu\k\l}R^{\mu\nu\k\l}\sim \frac{8(n+1)\al^2}{n(n-1)^2\e^2}.
\end{equation}
It should be noted that the Kretschmann scalar at the infinity behaves identically to corresponding value for the chargeless black hole \cite{Stetsko} and this fact is obvious because both types of black holes have asymptotically AdS-like behaviour at the infinity. To sum up the investigation of Kretschmann scalar we can conclude that its behaviour is rather typical for a black hole with the only singularity at the origin, coordinate singularities at the horizons and some asymptotic value at the infinity.

Now we can proceed to the investigation of the solutions of equations for  gravitational field in the case of negative $d^2$ ($d^2<0$). This situation takes place when the parameters $\e$ and $\ve$ have opposite signs.  The general formula (\ref{U_int}) is also valid in this case, but due to the negativity of $d^2$ the results of integration would be different. Not to introduce new notations here we denote $d^2=-|d|^2$ and $|\cdot|$ means the absolute value of the parameter. The results of integration again depend on the parity of $n$ and the fact whether $N_1$ and $N_2$ are integer or not. For the first integral in the relation (\ref{U_int}) we write:
\begin{equation}\label{i_1_neg}
\int \frac{r^{n+1}}{r^2-|d|^2}dr=\frac{|d|^n}{2}\ln\left|{\frac{r-|d|}{r+|d|}}\right|+\sum^{(n-1)/2}_{j=0}|d|^{2j}\frac{r^{n-2j}}{n-2j}
\end{equation}
for odd values of $n$ and
\begin{equation}\label{i_2_neg}
\int \frac{r^{n+1}}{r^2-|d|^2}dr=\frac{|d|^n}{2}\ln\left|\frac{r^2}{|d|^2}-1\right|+\sum^{n/2-1}_{j=0}|d|^{2j}\frac{r^{n-2j}}{n-2j}
\end{equation}
for even $n$. For noninteger values of $N_1$ and $N_2$ we can utilize the relation (\ref{i_3}) making the replacement $-d^2\rightarrow|d|^2$ and for integer values of $N_i$ it can calculated directly and written in terms of elementary functions or having the representation through the special functions (\ref{i_3}) and using some limit procedures, when parameters of the hypergeometric functions become integer.

Now we can write the expression for the metric function $U(r)$, namely for odd $n$ and $r>|d|$ we arrive at the following expression:
 \begin{eqnarray}\label{funct_odd_neg}
\nonumber U(r)=\ve-\frac{\mu}{r^{n-2}}-\frac{2\L}{n(n-1)}r^2+\frac{(\al+\L\e)^2}{2\al\e(n-1)}\left[\frac{|d|^n}{2r^{n-2}}\ln\left|{\frac{r-|d|}{r+|d|}}\right|+\sum^{(n-1)/2}_{j=0}|d|^{2j}\frac{r^{2(1-j)}}{n-2j}\right]\\\nonumber-2^p\frac{(2p-1)^2q^{2p}}{(2p-n)(n-1)}r^{\frac{2(3p-pn-1)}{2p-1}}+\frac{2^{p-1}(2p-1)^2(\al+\L\e)q^{2p}}{\al(n-1)(2p-n)}r^{2\left(1-\frac{p(n-1)}{2p-1}\right)}\times\\\nonumber{_{2}F_{1}}\left(1,\frac{p(n-1)}{2p-1}-\frac{n}{2};\frac{p(n-1)}{2p-1}-\frac{n}{2}+1;\frac{|d|^2}{r^2} \right)+\frac{2^{2p-3}(2p-1)^3\e q^{4p}}{\al(n-1)(4p-2pn-n)}r^{2\left(1-\frac{2p(n-1)}{2p-1}\right)}\times\\{_{2}F_{1}}\left(1,\frac{2p(n-1)}{2p-1}-\frac{n}{2};\frac{2p(n-1)}{2p-1}-\frac{n}{2}+1;\frac{|d|^2}{r^2} \right),
 \end{eqnarray}
 whereas for the even $n$ and $r>|d|$ we obtain:
 \begin{eqnarray}\label{funct_even_neg}
\nonumber U(r)=\ve-\frac{\mu}{r^{n-2}}-\frac{2\L}{n(n-1)}r^2+\frac{(\al+\L\e)^2}{2\al\e(n-1)}\left[\frac{|d|^n}{2r^{n-2}}\ln\left|{\frac{r^2}{|d|^2}}-1\right|+\sum^{n/2-1}_{j=0}|d|^{2j}\frac{r^{2(1-j)}}{n-2j}\right]\\\nonumber-2^p\frac{(2p-1)^2q^{2p}}{(2p-n)(n-1)}r^{\frac{2(3p-pn-1)}{2p-1}}+\frac{2^{p-1}(2p-1)^2(\al+\L\e)q^{2p}}{\al(n-1)(2p-n)}r^{2\left(1-\frac{p(n-1)}{2p-1}\right)}\times\\\nonumber{_{2}F_{1}}\left(1,\frac{p(n-1)}{2p-1}-\frac{n}{2};\frac{p(n-1)}{2p-1}-\frac{n}{2}+1;\frac{|d|^2}{r^2} \right)+\frac{2^{2p-3}(2p-1)^3\e q^{4p}}{\al(n-1)(4p-2pn-n)}r^{2\left(1-\frac{2p(n-1)}{2p-1}\right)}\times\\{_{2}F_{1}}\left(1,\frac{2p(n-1)}{2p-1}-\frac{n}{2};\frac{2p(n-1)}{2p-1}-\frac{n}{2}+1;\frac{|d|^2}{r^2} \right).
 \end{eqnarray}
 We remark that similarly to the previously considered case of positive $d^2$  another representation of hypergeometric function should be utilized when $r<|d|$. It should be noted that the given relations for the metric function have coordinate singularity at the point $r=d$ which is also the point  where the function $(\vp')^2$ becomes singular. It is easy to check that for the distances larger than $|d|$ the function $U(r)$ for both parities of $n$ ((\ref{funct_odd_neg}) and (\ref{funct_even_neg})) has a cosmological horizon in complete analogy with the chargeless case \cite{Stetsko}, thus it is difficult to treat the obtained solution as a black hole. 
 
 In the following we will consider thermodynamics of black holes represented by the metric functions (\ref{funct_odd}) and (\ref{funct_even}). Since the black hole's mass is extremely important for the thermodynamic we use the relations (\ref{funct_odd}) and (\ref{funct_even}) to express the mass parameter $\mu$ as a function of horizon radius $r_+$. As a result we obtain:
\begin{eqnarray}\label{mass_param_odd}
\nonumber\mu=\ve\left(1-\frac{(\al+\L\e)^2}{4\al^2}\right)r^{n-2}_{+}+\frac{(\al-\L\e)^2}{2n(n-1)\al\e}r^n_{+}+\frac{(\al+\L\e)^2}{2\al\e(n-1)}\left[(-1)^{\frac{n+1}{2}}d^n\arctan\left({\frac{r_+}{d}}\right)+\right.\\\left.\nonumber\sum^{(n-1)/2}_{j=2}(-1)^jd^{2j}\frac{r^{n-2j}_+}{n-2j}\right]-\frac{2^p(2p-1)^2q^{2p}}{(2p-n)(n-1)}r^{\frac{2p-n}{2p-1}}_{+}\left(1-\frac{(\al+\L\e)}{2\al}\times\right.\\\left.\nonumber{_{2}F_{1}}\left(1,\frac{p(n-1)}{2p-1}-\frac{n}{2};\frac{p(n-1)}{2p-1}-\frac{n}{2}+1;-\frac{d^2}{r^2_+}\right)\right)+\frac{2^{2p-3}(2p-1)^3\e q^{4p}}{\al(n-1)(4p-2pn-n)}\times\\r^{\frac{4p-2pn-n}{2p-1}}_{+}{_{2}F_{1}}\left(1,\frac{2p(n-1)}{2p-1}-\frac{n}{2};\frac{2p(n-1)}{2p-1}-\frac{n}{2}+1;-\frac{d^2}{r^2_+}\right)
\end{eqnarray}
for odd $n$ and for even $n$ we arrive at:
\begin{eqnarray}\label{mass_param_even}
\nonumber\mu=\ve\left(1-\frac{(\al+\L\e)^2}{4\al^2}\right)r^{n-2}_{+}+\frac{(\al-\L\e)^2}{2n(n-1)\al\e}r^n_{+}+\frac{(\al+\L\e)^2}{2\al\e(n-1)}\left[(-1)^{\frac{n}{2}}\frac{d^n}{2}\ln\left({\frac{r^2_+}{d^2}}+1\right)+\right.\\\left.\nonumber\sum^{n/2-1}_{j=2}(-1)^jd^{2j}\frac{r^{n-2j}_+}{n-2j}\right]-\frac{2^p(2p-1)^2q^{2p}}{(2p-n)(n-1)}r^{\frac{2p-n}{2p-1}}\left(1-\frac{(\al+\L\e)}{2\al}\times\right.\\\left.\nonumber{_{2}F_{1}}\left(1,\frac{p(n-1)}{2p-1}-\frac{n}{2};\frac{p(n-1)}{2p-1}-\frac{n}{2}+1;-\frac{d^2}{r^2_+}\right)\right)+\frac{2^{2p-3}(2p-1)^3\e q^{4p}}{\al(n-1)(4p-2pn-n)}\times\\r^{\frac{4p-2pn-n}{2p-1}}_{+}{_{2}F_{1}}\left(1,\frac{2p(n-1)}{2p-1}-\frac{n}{2};\frac{2p(n-1)}{2p-1}-\frac{n}{2}+1;-\frac{d^2}{r^2_+}\right)
\end{eqnarray}
\begin{figure}
\centerline{\includegraphics[scale=0.33,clip]{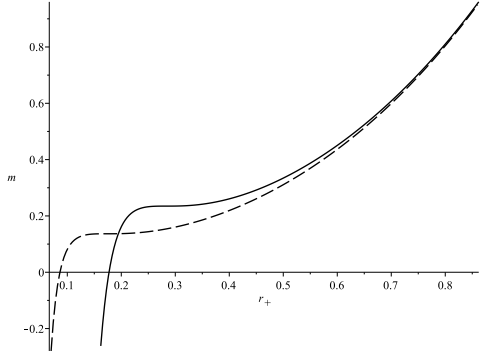}}
\caption{Mass parameter $\mu$ as a function of the horizon radius $r_+$ for linear ($p=1$, solid  curve) and conformal ($p=\frac{n+1}{4}$, dashed curve) types of gauge field. All the other parameters are equal for both cases, namely $n=4$, $\al=0.2$, $\e=0.4$, $\ve=1$, $\L=-2$, $q=0.2$.}\label{mass_f_graph}
\end{figure}
The dependence $\mu=\mu(r_+)$ is illustrated on the Fig.[\ref{mass_f_graph}]. One can see that the crucial difference between them is in range of small $r_+$, whereas for large $r_+$ the difference between them becomes negligibly small and this fact can be explained by the dominance of the $\sim r^n_+$ term, which has the same factor for all values of the parameter $p$.
 
\subsection{Field potential and black hole's charge}
Since we have obtained a charged black hole and we are going to obtain the first law of black hole thermodynamics in the following it is very important to have the relation for the total charge of the black hole and the evident form of the vector potential. As it was noted above, the potential has the only component $A_0(r)$ which can be recovered by integration of the relation (\ref{EM_field}), namely we can write:
\begin{equation}\label{int_pot}
A_t\equiv\psi=\psi_0-\int\frac{q}{r^{\frac{n-1}{2p-1}}}\sqrt{UW}dr
\end{equation}
where $\psi_0$ denotes a constant which is taken to impose some condition on the value of potential, namely in our case we suppose that $\psi_0$ is nonzero, taken in a such way that at the event horizon the the potential $\psi$ is equal to zero. It should be noted that we can choose the constant in other way, namely we can impose that at the infinity $\psi\rightarrow 0$, but our choice is a bit convenient when we consider black hole thermodynamics. Having substituted the relation (\ref{UW_prod}) into (\ref{int_pot}) and performing integration we can write:
\begin{eqnarray}\label{pot_gen_i}
\nonumber\psi(r)=\psi_0+\frac{2p-1}{n-2p}qr^{\frac{2p-n}{2p-1}}+\frac{(2p-1)(\al+\L\e)q}{2(2p-n)\al}r^{\frac{2p-n}{2p-1}}{_{2}F_{1}}\left(1,\frac{n-1}{2(2p-1)}-\frac{1}{2};\frac{n-1}{2(2p-1)}+\frac{1}{2};-\frac{d^2}{r^2}\right)\\
+\frac{2^{p-2}(2p-1)\e q^{2p+1}}{(4p-2pn-n)\al}r^{1-\frac{(2p+1)(n-1)}{2p-1}}{_{2}F_{1}}\left(1,\frac{(2p+1)(n-1)}{2(2p-1)}-\frac{1}{2};\frac{(2p+1)(n-1)}{2(2p-1)}+\frac{1}{2};-\frac{d^2}{r^2}\right)
\end{eqnarray}
and here it is assumed that $\frac{2p(n-1)}{2p-1}$ is noninteger. It should also be remarked that the written hypergeometric functions are valid for large distances ($r>d$) and they tend to one when $r\rightarrow\infty$, so at large distances the potential $\psi$ demonstrate power-law dependence. Similarly to the metric function $U(r)$  for small distances ($r<d$) another representation for the hypergeometric functions should be used, but again in the limit $r\rightarrow 0$ the potential shows power-law dependence. To obtain finite value of the potential at the infinity one should impose that $\frac{2p-n}{2p-1}<0$ which lead to restriction on the value of parameter of nonlinearity $p$, namely we obtain:
\begin{equation}\label{cond_p}
\frac{1}{2}<p<\frac{n}{2}.
\end{equation}
It should be noted that written above condition for the parameter $p$ is only related to electromagnetic part of the action and there are no imprints of gravity part. We also remark that some conditions on the parameter of nonlinearity was obtained in the paper \cite{Kord_PRD15} where a black hole in Einstein-dilaton-nonlinear Maxwell theory was investigated. The quite general conclusion one can make here is that the parameter of nonlinearity cannot be chosen completely arbitrary, in any kind of theory one has some conditions which are imposed on the parameter of nonlinearity and those conditions are taken to obtain reasonable  behaviour of electric potential and possibly some other functions.

In case of linear field $p=1$ (it satisfies the condition (\ref{cond_p})) the gauge potential (\ref{int_pot}) can be written in terms of elementary functions. We write them in the separate form for odd and even $n$ respectively. Namely, for odd $n$ one arrives at:
\begin{eqnarray}\label{pot_lin_odd}
\nonumber\psi(r)=\psi_0+\frac{q}{n-2}r^{2-n}+\frac{(\al+\L\e)q}{2\al}\left[\sum^{(n-5)/2}_{j=0}(-1)^j\frac{r^{2j-n+4}}{d^{2(j+1)}(2j-n+4)}+\frac{(-1)^{\frac{n-3}{2}}}{d^{n-2}}\arctan\left(\frac{r}{d}\right)\right]\\+\frac{\e q^3}{2\al}\left[\sum^{(3n-7)/2}_{j=0}(-1)^j\frac{r^{2j-3n+6}}{d^{2(j+1)}(2j-3n+6)}+\frac{(-1)^{\frac{3n-5}{2}}}{d^{3n-4}}\arctan\left(\frac{r}{d}\right)\right],
\end{eqnarray} 
and for even $n$ the potential is as follows:
\begin{eqnarray}\label{pot_lin_even}
\nonumber\psi(r)=\psi_0+\frac{q}{n-2}r^{2-n}+\frac{(\al+\L\e)q}{2\al}\left[\sum^{(n-6)/2}_{j=0}(-1)^j\frac{r^{2j-n+4}}{d^{2(j+1)}(2j-n+4)}+\frac{(-1)^{\frac{n}{2}+1}}{2d^{n-2}}\ln\left(1+\frac{d^2}{r^2}\right)\right]\\+\frac{\e q^3}{2\al}\left[\sum^{(3n-8)/2}_{j=0}(-1)^j\frac{r^{2j-3n+6}}{d^{2(j+1)}(2j-3n+6)}+\frac{(-1)^{\frac{3n}{2}}}{2d^{3n-4}}\ln\left(1+\frac{d^2}{r^2}\right)\right].
\end{eqnarray} 
It should be noted that for large $r$ the potential for both parities of $n$ (relations (\ref{pot_lin_odd}) and (\ref{pot_lin_even})) has reasonable behavior, because it contains the terms of inverse proportionality ($\sim\frac{1}{r^{m}}$, $m>0$), and it means that at the infinity the potential is finite.

We also note that for the case when $p=\frac{n+1}{4}$ (conformal case) the gauge potential can also be written in terms of elementary functions. So for this type of nonlinearity we can write for odd $n$:
\begin{eqnarray}\label{pot_conf_odd}
\nonumber\psi(r)=\psi_0+\frac{q}{r}+\frac{(\al+\L\e)q}{2\al d}\arctan\left(\frac{r}{d}\right)+2^{\frac{n-3}{4}}\frac{(n-1)\e q^{\frac{n+3}{2}}}{4\al}\times\\\left[\sum^{(n+1)/2}_{j=0}\frac{(-1)^jr^{2j-2-n}}{d^{2(j+1)}(2j-2-n)}+\frac{(-1)^{\frac{n+3}{2}}}{d^{n+4}}\arctan\left(\frac{r}{d}\right)\right];
\end{eqnarray}
and for even $n$ we arrive at:
\begin{eqnarray}\label{pot_conf_even}
\nonumber\psi(r)=\psi_0+\frac{q}{r}+\frac{(\al+\L\e)q}{2\al d}\arctan\left(\frac{r}{d}\right)+2^{\frac{n-3}{4}}\frac{(n-1)\e q^{\frac{n+3}{2}}}{4\al}\times\\\left[\sum^{n/2}_{j=0}\frac{(-1)^jr^{2j-2-n}}{d^{2(j+1)}(2j-2-n)}+\frac{(-1)^{\frac{n}{2}}}{2d^{n+4}}\ln\left(1+\frac{d^2}{r^2}\right)\right];
\end{eqnarray}
Having compared the written above relations for potential in case of conformal nonlinearity with the case of linear field one can conclude that    the conformal case has the terms that go slower to zero at the infinity than in linear case, so in the considered above linear case the potential goes faster to its limit values than in conformal case.

Apart of the gauge potential we have just considered, the other important quantity is the charge of the black hole. To obtain it we utilize Gauss law which in nonlinear case can be represented in the form:
\begin{equation}\label{Gauss_law}
Q=\frac{1}{4\pi}\int_{\Sigma}(-F_{\mu\nu}F^{\mu\nu})^{p-1}*F,
\end{equation}
where $*F$ denotes a Hodge dual of electromagnetic field form and integral is taken over $n-1$-dimensional closed spacelike hypersurface $\Sigma$. Having calculated the latter integral we arrive at:
\begin{equation}\label{total_charge}
Q=\frac{\omega_{n-1}}{4\pi}2^{p-1}q^{2p-1}
\end{equation}
and here $\omega_{n-1}$ denotes the volume (or surface area) of ``unit'' hypersurface of constant curvature (it reduces to the surface area of unit sphere in case of spherical symmetry). Now we can define total electric charge of the black hole per unit volume $\omega_{n-1}$ as follows:
\begin{equation}
\bar{Q}=\frac{1}{\omega_{n-1}}Q=\frac{1}{4\pi}2^{p-1}q^{2p-1}
\end{equation}
The electric potential that an observer can measure at infinity with respect to the horizon can be defined in the form:
\begin{equation}\label{phi_pot}
\Phi_q=A_{\mu}\chi^{\mu}\Big|_{+\infty}-A_{\mu}\chi^{\mu}\Big|_{r_+}
\end{equation}
where $\chi^{\mu}$ is a Killing vector, null on the horizon. If one considers linear gauge theory ($p=1$) the Killing vector might be the time translation vector $\chi^{\mu}=\partial/\partial t$, but for $p\neq 1$ to obtain consistent theory one should make other choice for the Killing vector, namely it was shown that for power-law field theory one should take : $\chi^{\mu}=p\partial/\partial t$ \cite{Kord_PRD15} and as a result the electric potential of the black hole takes the form:
\begin{equation}\label{phi_val}
\Phi_q=p\psi_0.
\end{equation}
One can see that the electric potential the observer can measure is completely defined by the integration constant $\psi_0$, which provides the  condition that the gauge potential $A_{\mu}$  is equal to zero at the horizon and the parameter of nonlinearity $p$.  It should be pointed out that one can impose that $A_{\mu}=0$ at the infinity but the electric potential $\Phi_{q}$ defined by the relation (\ref{phi_pot}) would be completely identical to the expression (\ref{phi_val}).

\section{Thermodynamics of the black hole}
\subsection{Black hole temperature}
To obtain black hole\rq{}s thermodynamics we start from calculation of its temperature. The temperature is supposed to be well defined quantity in quite general setting due to its strict geometrical basis \cite{Gibbons_PRD77}. To calculate the temperature we use the so called surface gravity which is defined as follows:
\begin{equation}\label{surf_grav}
\kappa^2=-\frac{1}{2}\nabla_{a}\chi_b\nabla^{a}\chi^{b},
\end{equation}
where $\chi^{\mu}$ is a Killing vector field which is null on the horizon. Because we consider the static metric we can chose $\chi^{\nu}=\partial/\partial t$. As a result we can write relation for black hole\rq{}s temperature:
\begin{equation}\label{BH_temp}
T=\frac{\k}{2\pi}=\frac{1}{4\pi}\frac{U\rq{}(r_+)}{\sqrt{U(r_+)W(r_+)}}
\end{equation}  
and here $r_+$ similarly as in the previous section denotes the   event horizon\rq{}s radius. It is well understood that the evident form for black hole's temperature will depend on the parity of $n$ and the fact whether the mentioned before numbers $N_1$ and $N_2$ are integers or not. Under the assumption that the parameters $N_1$ and $N_2$ are noninteger we can write the relation for temperature as follows:
\begin{eqnarray}\label{temp_gen}
\nonumber T=\frac{1}{4\pi}\frac{2\al r^2_{+}+\ve\e(n-1)(n-2)}{(\al-\L\e)r^2_{+}+\ve\e(n-1)(n-2)-2^{p-1}(2p-1)\e q^{2p}r^{2\left(1-\frac{p(n-1)}{2p-1}\right)}_+}\left(\frac{(\al-\L\e)^2}{2(n-1)\al\e}r_{+}+\right.\\\nonumber\left.\ve\left(1-\frac{(\al+\L\e)^2}{4\al^2}\right)\frac{(n-2)}{r_+}+\frac{(\al+\L\e)^2}{2(n-1)\al\e}\left[(-1)^{\frac{(n+1-\sigma)}{2}}\frac{d^{n+1-\sigma}}{r^{n-2-\sigma}_{+}(r^2_{+}+d^2)}+\sum^{(n-1-\sigma)/2}_{j=2}(-1)^jd^{2j}r^{1-2j}_{+}\right]\right.\\\nonumber\left.-\frac{2^p(2p-1)q^{2p}}{n-1}r^{1-\frac{2p(n-1)}{2p-1}}_{+}\left[1-\frac{(\al+\L\e)}{2\al}\left({_{2}F_{1}}\left(1,\frac{p(n-1)}{2p-1}-\frac{n}{2};\frac{p(n-1)}{2p-1}-\frac{n}{2}+1;-\frac{d^2}{r^2_{+}} \right)-\right.\right.\right.\\\nonumber\left.\left.\left.\frac{2(2p-1)d^2}{(n+2p-2)r^2_{+}}{_{2}F_{1}}\left(2,\frac{p(n-1)}{2p-1}-\frac{n}{2}+1;\frac{p(n-1)}{2p-1}-\frac{n}{2}+2;-\frac{d^2}{r^2_{+}} \right)\right)\right]+\frac{2^{2p-3}(2p-1)^2\e q^{4p}}{\al(n-1)}\times\right.\\\nonumber\left.r^{1-\frac{4p(n-1)}{2p-1}}_{+}\left[{_{2}F_{1}}\left(1,\frac{2p(n-1)}{2p-1}-\frac{n}{2};\frac{2p(n-1)}{2p-1}-\frac{n}{2}+1;-\frac{d^2}{r^2_{+}} \right)-\frac{2(2p-1)d^2}{(2pn+n-2)r^2_{+}}\times\right.\right.\\\left.\left.{_{2}F_{1}}\left(2,\frac{2p(n-1)}{2p-1}-\frac{n}{2}+1;\frac{2p(n-1)}{2p-1}-\frac{n}{2}+2;-\frac{d^2}{r^2_{+}} \right)\right]\right)
\end{eqnarray} 
and here $\sigma=0$ when $n$ is odd and $\sigma=1$ for even $n$. It is worth noting that latter relation represents the temperature as a function of the horizon radius $r_+$, for $r_+>d$, if $r_+<d$ the representation of integrals given by the relation (\ref{i_4}) should be utilized in the metric function (\ref{U_int}) instead of the relation (\ref{i_3}) which has brought  us to the written above relation (\ref{temp_gen}). The behaviour of the function $T=T(r_+)$ is rather complicated, but nevertheless looking at the relation (\ref{temp_gen}) some conclusions about the behaviour of this function can be made. The most important point here that should be paid attention to is the fact that for large $r_+$ the temperature grows up almost linearly due to corresponding linear term and the fact that all the other terms are inversely proportional to some power of $r_+$. The hypergeometric functions in the relation (\ref{temp_gen}) go to one when $r_{+}\rightarrow\infty$ and due to the presence of factors $r^{1-2p(n-1)/(2p-1)}_+$, $r^{1-4p(n-1)/(2p-1)}_+$ and $r^{-2}_+$ the contribution of the terms related to electromagnetic field in the temperature (\ref{temp_gen}) becomes negligibly small when $r_+\rightarrow\infty$. As it has just been noted, to obtain general relation for the temperature when $r_+<d$ one should use the relation (\ref{i_4}) and the resulting hypergeometric functions one arrives at will go to one when $r_+\rightarrow 0$, but the terms with hypergeometric functions contain the factors of inverse proportionality $\sim 1/r^{\gamma}_+$  ($\gamma>0$), thus the black hole temperature would have power-law singularity when $r_+\rightarrow 0$. So having relied on the mentioned above facts we can conclude that the temperature might have nonmonotonous behaviour which gives rise to the consequence that in this case we have a phase transition. It is also worth being remarked that a Hawking-Page phase transition usually takes place in AdS-like cases, due to this fact  the phase transition for our black hole supposedly would be of that type. 

For the particular case of linear field ($p=1$) we can write the temperature in the following form:
\begin{eqnarray}\label{temp_linear}
\nonumber T=\frac{1}{4\pi}\frac{2\al r^2_{+}+\ve\e(n-1)(n-2)}{(\al-\L\e)r^2_{+}+\ve\e(n-1)(n-2)-\e q^{2}r^{2(2-n)}_+}\left(\frac{(\al-\L\e)^2}{2(n-1)\al\e}r_{+}+\ve\left(1-\frac{(\al+\L\e)^2}{4\al^2}\right)\times\right.\\\nonumber\left.\frac{(n-2)}{r_+}+\frac{(\al+\L\e)^2}{2(n-1)\al\e}\left[\frac{(-1)^{\frac{(n+1-\sigma)}{2}}d^{n+1-\sigma}}{r^{n-2-\sigma}_{+}(r^2_{+}+d^2)}+\sum^{\frac{n-1-\sigma}{2}}_{j=2}(-1)^jd^{2j}r^{1-2j}_{+}\right]-\frac{2q^2}{n-1}r^{3-2n}_{+}+\frac{(\al+\L\e)q^2}{\al(n-1)}\times\right.\\\left.\left[\frac{(-1)^{\frac{(n-3-\sigma)}{2}}d^{3+\sigma-n}}{r^{n-2+\sigma}_{+}(r^2_{+}+d^2)}+\sum^{\frac{n-5-\sigma}{2}}_{j=0}(-1)^j\frac{r^{2j+5-2n}_{+}}{d^{2(j+1)}}\right]+\frac{\e q^4}{2(n-1)\al}\left[\frac{(-1)^{\frac{(3n-5-\sigma)}{2}}d^{5+\sigma-3n}}{r^{n-2+\sigma}_{+}(r^2_{+}+d^2)}+\sum^{\frac{3n-7-\sigma}{2}}_{j=0}(-1)^j\frac{r^{2j+7-4n}_{+}}{d^{2(j+1)}}\right]\right)
\end{eqnarray}
The written expression for the temperature in linear case ($p=1$) inherits the main features that we described for the nonlinear case ($p\neq 1$). In particular, it has almost linear increase for large radii, distorted by a weak nonlinearities which become negligibly small for large $r_+$. If the raduis of horizon $r_+$ decreases and for some value of $r_+$ it might turn negative, but this negativity would mean that the black holes with the other fixed parameters does not exist. When $r_+\rightarrow 0$ the temperature remains negative and goes down to infinitely low values and this decrease is mainly driven by the term $\sim -1/{r^{2n-3}}$.

For the other interesting particular case, namely when $p=(n+1)/4$ we have the following expression for temperature:
\begin{eqnarray}\label{temp_conf}
\nonumber T=\frac{1}{4\pi}\frac{2\al r^2_{+}+\ve\e(n-1)(n-2)}{(\al-\L\e)r^2_{+}+\ve\e(n-1)(n-2)-2^{\frac{n-7}{4}}(n-1)\e q^{\frac{n+1}{2}}r^{1-n}_{+}}\left(\frac{(\al-\L\e)^2}{2\al\e(n-1)}r_{+}+\right.\\\nonumber\left.\ve\left(1-\frac{(\al+\L\e)^2}{4\al^2}\right)\frac{(n-2)}{r_+}-2^{\frac{n-3}{4}}\frac{q^{\frac{n+1}{2}}}{r^{n}_{+}}+\frac{2^{\frac{n-7}{4}}(\al+\L\e)q^{\frac{n+1}{2}}}{\al(r^2_{+}+d^2)r^{n-2}_{+}}+\frac{(\al+\L\e)^2}{2\al\e(n-1)}\left[\sum^{\frac{n-1-\sigma}{2}}_{j=2}(-1)^jd^{2j}r^{1-2j}_{+}\right.\right.\\\left.\left.+\frac{(-1)^{\frac{n+1-\sigma}{2}}d^{n+1-\sigma}}{r^{n-2-\sigma}_{+}(r^2_{+}+d^2)}\right]+2^{\frac{n-9}{2}}\frac{(n-1)\e q^{n+1}}{\al}\left[\sum^{\frac{n-1-\sigma}{2}}_{j=0}(-1)^j\frac{r^{2(j-n)+1}_{+}}{d^{2(j+1)}}+\frac{(-1)^{\frac{n+1-\sigma}{2}}}{d^{n+1-\sigma}r^{n-2+\sigma}_{+}(r^2_{+}+d^2)}\right]\right).
\end{eqnarray} 
The given expression shows that similarly to the previously considered case for large $r_+$ the temperature (\ref{temp_conf}) grows up almost linearly with increasing of $r_+$, while for infinitely low values of the horizon radius $r_+$ ($r_+\rightarrow 0$) the behaviour of the temperature is defined by the leading term $\sim -1/r^{n}_{+}$, and in comparison with the previously examined linear case in the conformal case ($p=(n+1)/4$) the temperature decreases slowly. It should be stressed here that the mentioned above ranges of horizon radii $r_+$ when the temperature of the black holes turns to be negative are unphysical and we refer them just to have more complete information about the function $T=T(r_+)$.  
\begin{figure}
\centerline{\includegraphics[scale=0.33,clip]{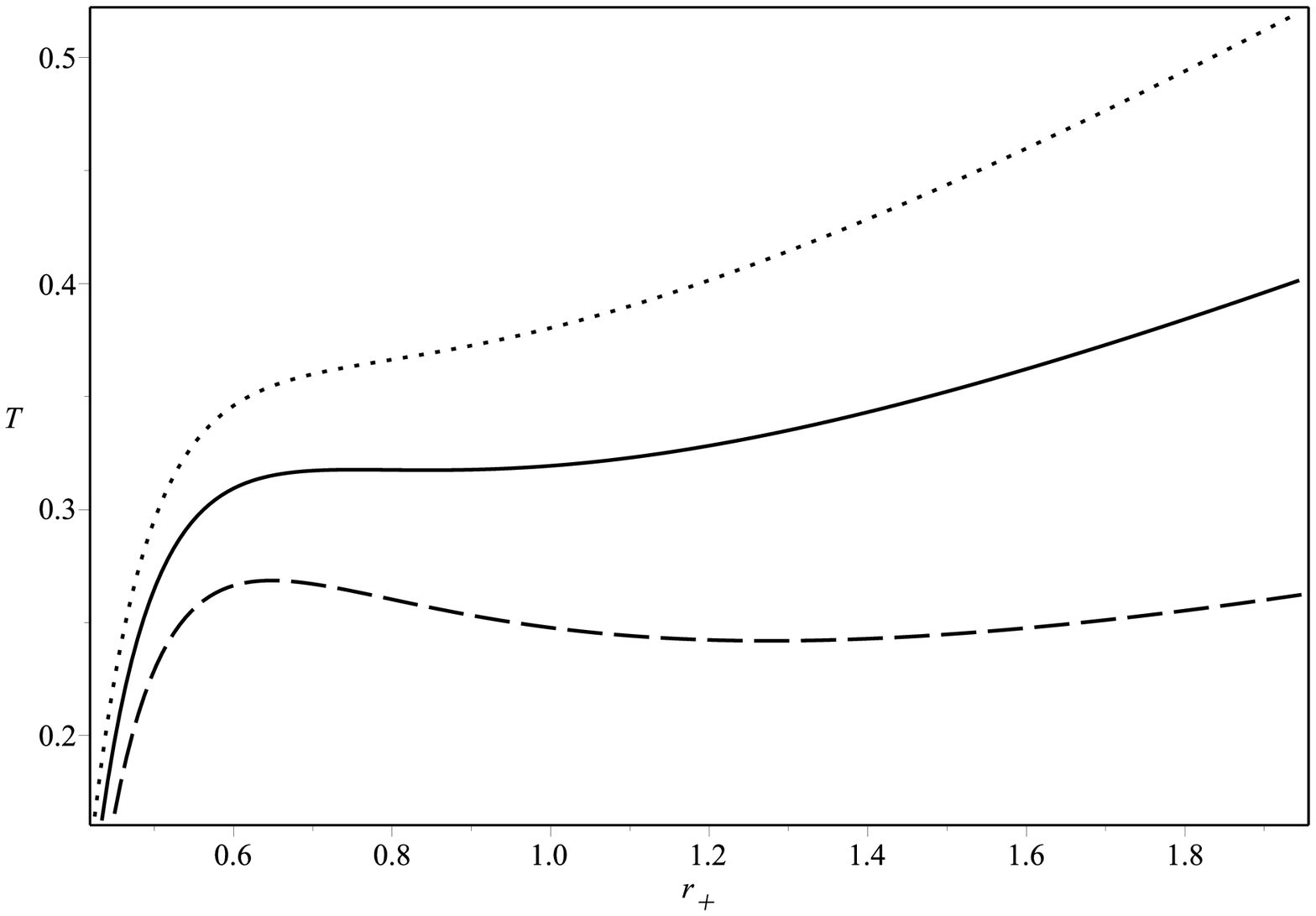}\includegraphics[scale=0.303,clip]{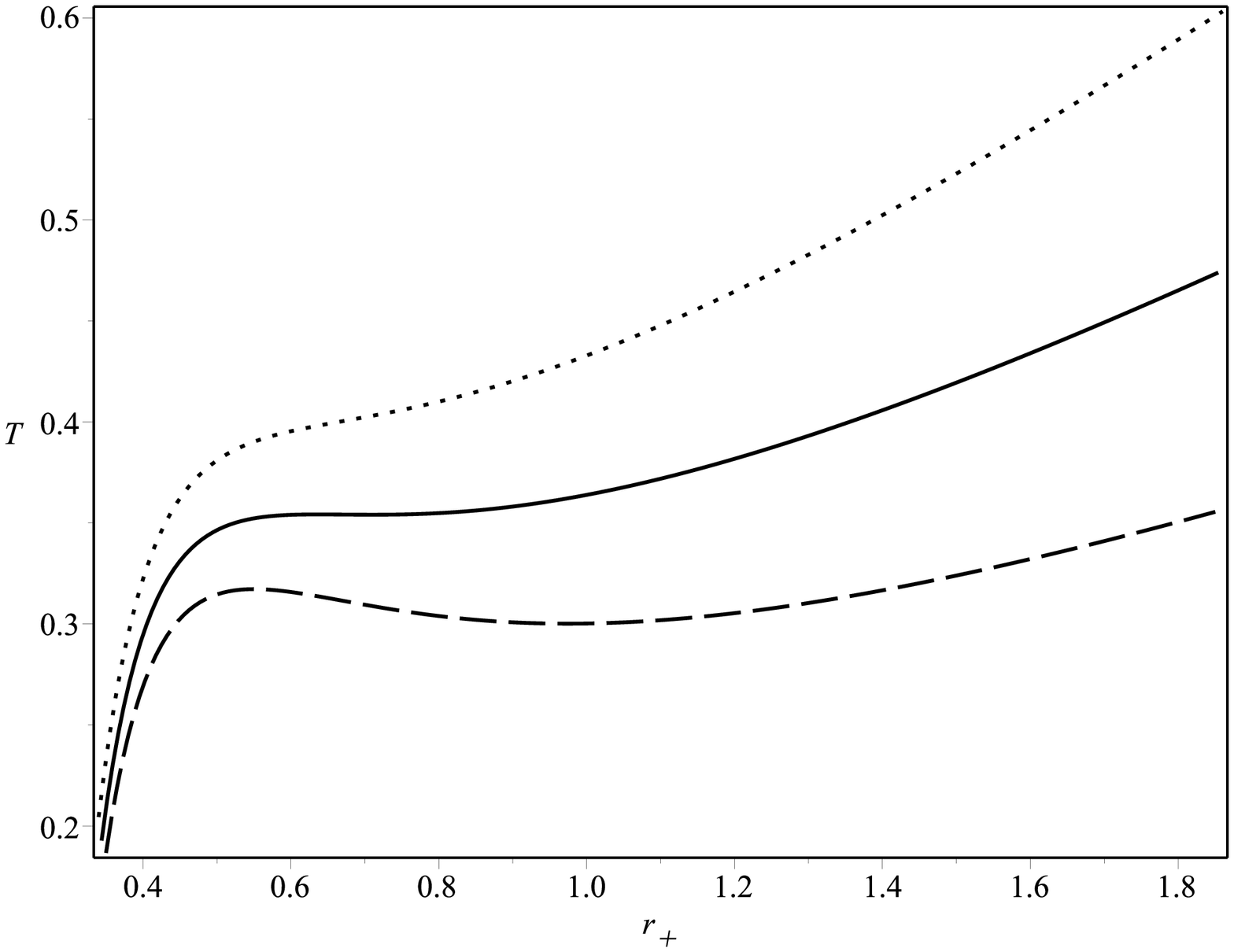}}
\caption{Black hole temperature $T$ as a function of the horizon radius for linear $p=1$ (the left graph) and conformal $p=\frac{n+1}{4}$ (the right one). For both graphs the solid curves represent almost critical value of the cosmological constant ($\L_c$), the dotted curves correspond to the values of $\L$ above the critical one, the dashed curves correspond to the values below the critical. For both graphs some parameters are the same, namely $n=4$, $\al=0.2$, $\e=0.4$, $\ve=1$ and $q=0.4$. For the left graph correspondence of curves is as follows (from the lowest the highest pressures): $\L=-3$, $\L_c=-5.7$, $\L=-8$ and for the right graph we have the following choice of the parameters: $\L=-5$, $\L_c=-7.4$, $\L=-10$.}\label{Temp_graph}
\end{figure}

The Fig.[\ref{Temp_graph}] shows the dependence of the temperature $T$ on the horizon radius $r_+$ for linear as well as conformal cases. Those dependences have some common features, namely for some values of cosmological constant greater in absolute value than some fixed (critical) one the temperature goes up monotonically with increasing of the radius of horizon $r_+$. As it will be shown below that the thermodynamic pressure we are going to introduce is proportional to the absolute value of the cosmological constant \cite{Kastor_CQG09} and as a consequence the increasing of the module of $\L$ means the increasing of the pressure. Under the assumption about thermodynamic pressure the Fig.[\ref{Temp_graph}] has direct analogy with the behaviour of $TV$-diagram of a substance described by Van der Waals equation of state which is applicable to real systems in condensed matter physics such as liquids or gases \cite{Kubiznak_JHEP12,Gunasekaran_JHEP12}.

\subsection{Wald procedure and entropy}
Consistent approach which allows one to obtain conserved charges of a black hole was developed by Wald \cite{Wald_PRD93,Iyer_PRD94}. Having used the derived conserved quantities one can develop black hole's thermodynamics. It should be noted that Wald formalism is applicable to general diffeomorphism-invariant theories even those containing the derivatives of higher order. Wald formalism was applied to numerous  examples of black holes, in particular, Einstein-scalar theory \cite{Liu_PLB14,Lu_JHEP15}, Einstein-Proca theory \cite{Liu_JHEP14}, Einstein-Yang-Mills theory \cite{Fan_JHEP15}, gravity with quadratic-curvature invariants \cite{Fan_PRD15}, Lifshitz black hole \cite{Liu_JHEP14_2}, black holes in Horndeski theory \cite{Feng_JHEP15,Feng_PRD16,Stetsko}.

It was argued \cite{Feng_JHEP15,Feng_PRD16} that the Wald relation for entropy should be corrected due to singular behaviour of the scalar potential at the horizon. To obtain the first law of black hole and relation for entropy the general Wald procedure was used \cite{Feng_JHEP15,Feng_PRD16}. The resulting expression for entropy differs from the well-known relation in General Relativity. In our work we also use the Wald procedure and before its application we briefly review the keypoints of this formalism.  Having supposed that we have some diffeomorphism-invariant theory described by a Lagrangian function ${\cal L}$, we perform the variation of the Lagrangian which can be written in the following form:
\begin{equation}
\delta{\cal L}={\rm e.o.m.}+\sqrt{-g}\nabla_{\nu}J^{\nu}
\end{equation}
and here ${\rm e.o.m.}$ represents the terms which give the equations of motion and $J^{\nu}$ is the surface ``current'' and the corresponding ``current'' $1$-form can be constructed $J_{(1)}=J_{\nu}dx^{\nu}$. This surface term allows to construct Hodge dual form $\Theta_{(n)}=*J_{(1)}$. Having chosen a vector of infinitesimal diffeomorphism $\delta x^{\mu}=\xi^{\mu}$ one constructs a form:
\begin{equation}
J_{(n)}=\Theta_{(n)}-i_{\xi}*{\cal L}=e.o.m.-d*J_{(2)}
\end{equation}
where $i_{\xi}*{\cal L}$ is the contraction of the vector field $\xi^{\mu}$ with the form $*{\cal L}$ (here $\cal L$ denotes the Lagrangian of the considered system). It should be noted that when the equations  of motion are satisfied (on-shell condition) one can consider $n-1$-form  $Q_{(n-1)}=*J_{(2)}$ for which $J_{(n)}=dQ_{(n-1)}$. To obtain some thermodynamic relations the vector field $\xi^{\mu}$ should be taken to be a Killing vector field that is null at the horizon. According to Wald approach the variation of the Hamiltonian can be represented in the form: 
\begin{equation}
\delta{\cal H}=\frac{1}{16\pi}\left(\delta \int_{c}J_{(n)}-\int_{c}d(i_{\xi}\Theta_{(n)})\right)=\frac{1}{16\pi}\int_{\Sigma_{(n-1)}}(\delta Q-i_{\xi}\Theta_{(n)})
\end{equation}
and here $c$ is a $n$--dimensional Cauchy surface and $\Sigma_{(n-1)}$ denotes its $n-1$-dimensional boundary with two components, namely one on the infinity and the other is on the event horizon. The first law of black hole thermodynamics follows from the relation:
\begin{equation}
\delta{\cal H}_{\infty}=\delta{\cal H}_{+}
\end{equation}
where indices of the variation of the Hamiltonian, namely $\infty$ and $+$ mean that these variations are taken at the infinity and at the event horizon respectively. It is clear that the first law of thermodynamics can be constructed if the variation $\delta{\cal H}$ is a well behaved function in the domain in between two mentioned above boundaries.

Now we can proceed to the direct calculations of the variation of the Hamiltonian. It should be noted that the Wald method in Horndeski theory was examined in \cite{Feng_JHEP15,Stetsko} for a chargeless black hole and in charged case it was considered in \cite{Feng_PRD16}.  We can write:
\begin{eqnarray}\label{variat_min}
\nonumber(\delta Q-i_{\xi}\Theta)_{min}=r\sqrt{UW}\left(\frac{(n-1)}{rW^2}\delta W+\frac{2p(2p-1)2^{p-1}}{(UW)^p}\left(\psi'\right)^{2(p-1)}\times\right.\\\left.\psi\left(\psi'\left(\frac{\delta W}{W}+\frac{\delta U}{U}\right)-2\delta\psi'\right)-\frac{\al\vp'}{W}\delta\vp\right)\Omega_{(n-1)},
\end{eqnarray}
\begin{eqnarray}\label{variat_nm}
\nonumber(\delta Q-i_{\xi}\Theta)_{nm}=\frac{\eta(n-1)}{2}r^{n-2}\sqrt{\frac{U}{W}}\left(\frac{(\vp')^2}{2W^2}\delta W-\delta\left(\frac{(\vp')^2}{W}\right)+\frac{2\al r}{(n-1)\e}\vp'\delta\vp\right)\Omega_{(n-1)},
\end{eqnarray}
where the relations (\ref{variat_min}) and (\ref{variat_nm}) are related to the minimally and nonminimally coupled parts of the action respectively. Their difference takes the form:
\begin{eqnarray}\label{variat_total}
\nonumber(\delta Q-i_{\xi}\Theta)_{tot}=r^{n-1}\sqrt{UW}\left(\frac{(n-1)}{rW^2}\left(1+\frac{\e}{4}\frac{(\vp')^2}{W}\right)\delta W+\frac{2p(2p-1)2^{p-1}}{(UW)^p}\left(\psi'\right)^{2(p-1)}\times\right.\\\left.\psi\left(\psi'\left(\frac{\delta W}{W}+\frac{\delta U}{U}\right)-2\delta\psi'\right)-\frac{\e(n-1)}{2rW}\delta\left(\frac{(\vp')^2}{W}\right)\right)\Omega_{(n-1)}.
\end{eqnarray}
It should be pointed out that if $p=1$ the latter relation gets transformed into the corresponding relation obtained for a linear field \cite{Feng_PRD16}. The important point we should pay attention to is the fact that for a chargeless black hole the latter term in the total variation (\ref{variat_total}), namely $\delta((\vp')^2/W)$ is equal to zero identically for all $r$, whereas for the charged linear case $p=1$ it equals to zero at infinity because of the decay of electromagnetic filed at infinity, at the same time the variation might not be equal to zero at  the horizon but it has the metric function $1/W$ as a factor and as a consequence the last term in the total variation also disappears on the horizon. In our case of nonlinear field we can write the this last term of the variation (\ref{variat_total}) at infinity in the following form:
\begin{equation}
r^{n-2}\frac{\sqrt{UW}}{W}\delta\left(\frac{(\vp')^2}{W}\right)\Big|_{\infty}=r^{n-2}\frac{U}{\sqrt{UW}}\delta\left(\frac{(\vp')^2}{W}\right)\Big|_{\infty}\longrightarrow\sim r^{n-\frac{2p(n-1)}{2p-1}}.
\end{equation}
As it is easy to see that for providing the disappearance of the given variation one has to impose some conditions on the parameter $p$, namely we get completely the same condition as the previously derived (\ref{cond_p}). This fact is an additional evidence that the parameter $p$ can not be taken arbitrary, but it should be chosen in the  interval (\ref{cond_p}) to obtain reasonable behaviour of the electric potential at infinity and as we have just seen to derive the  first law of black hole thermodynamics. At the horizon this variation is equal to zero identically as it was for linear case. Finally, we can write total variation at the infinity in the form:
\begin{equation}\label{var_inf}
(\delta Q-i_{\xi}\Theta)_{tot}=(n-1)\delta\mu-4p(2p-1)2^{p-1}q^{2(p-1)}\psi_{0}\delta q.
\end{equation}
Taking into account the relation for the total charge (\ref{total_charge}) and potential $\Phi_q$ (\ref{phi_val}) and performing integration over angular variables we can write the variation of total Hamiltonian at the infinity in the form:
\begin{equation}\label{var_H_inf}
{\cal H}_{\infty}=\delta M-\Phi_{q}\delta Q,
\end{equation}
where $\delta M$ is the variation of black hole's mass which can be defined as follows:
\begin{equation}\label{bh_mass}
M=\frac{(n-1)\omega_{n-1}}{16\pi}\mu.
\end{equation}
The obtained expression for the black hole's mass (\ref{bh_mass}) does not depend anyhow on the nonminimal coupling of scalar field $\vp$ with gravity and has the same form form as for Schwarzchild-AdS (or RN-AdS) black hole. Variating the Hamiltonian at the horizon we can write:
\begin{equation}\label{TD_diff}
\delta {\cal H}_{+}=\frac{(n-1)\omega_{n-1}}{16\pi}U\rq{}(r_+)r^{n-2}_{+}\delta r_{+}=\sqrt{U(r_+)W(r_+)}T\delta\left(\frac{{\cal A}}{4}\right)=\left(1+\frac{\e}{4}\frac{(\vp\rq{})^2}{W}\Big|_{r_+}\right)T\delta\left(\frac{{\cal A}}{4}\right).
\end{equation}
where ${\cal A}=\omega_{n-1}r^{n-1}_+$ is the horizon area of the black hole. One can see that it is not possible to represent the obtained variation in the form $T\delta S$ which is common in the standard General Relativity.  Now following the procedure proposed in \cite{Feng_PRD16} and developed in the work \cite{Stetsko} we can rewrite the latter relation as follows:
\begin{equation}\label{var_horizon}
{\cal H}_{+}=T\delta S+\Phi^{+}_{\vp}\delta Q^{+}_{\vp},
\end{equation}
where $S$ is entropy of the black hole, $Q^{+}_{\vp}$ is the so called scalar ``charge'' and corresponding to it potential $\Phi^{+}_{\vp}$  can be defined in the form:
\begin{equation}\label{entropy}
S=\left(1+\frac{\e}{4}\frac{(\vp\rq{})^2}{W}\Big|_{r_+}\right)\frac{{\cal A}}{4},
\end{equation}
\begin{equation}\label{sc_pot}
Q^{+}_{\vp}=\omega_{n-1}\sqrt{1+\frac{\e}{4}\frac{(\vp\rq{})^2}{W}\Big|_{r_+}}, \quad \Phi^{+}_{\vp}=-\frac{{\cal A}T}{2\omega_{n-1}}\sqrt{1+\frac{\e}{4}\frac{(\vp\rq{})^2}{W}\Big|_{r_+}}.
\end{equation}
We note that the chosen form for the scalar ``charge'' $Q^{+}_{\vp}$ and conjugate potential $\Psi^{+}_{\vp}$ is not unique, in particular in the work \cite{Feng_PRD16} different expressions were utilized. It can be verified easily that the chosen above forms for the scalar potential and its conjugate value (\ref{sc_pot}) allows one to obtain the following important relation:  
\begin{equation}\label{rel_entr_pot}
\Phi^{+}_{\vp}Q^{+}_{\vp}=-\frac{{\cal A}T}{2}\left(1+\frac{\e}{4}\frac{(\vp\rq{})^2}{W}\Big|_{r_+}\right)=-2TS
\end{equation}

Having equated the right hand sides of the relations (\ref{var_H_inf}) and (\ref{var_horizon}) one arrives at:
\begin{equation}\label{first_law}
\delta M=T\delta S+\Phi^{+}_{\vp}\delta Q^{+}_{\vp}+\Phi_{q}\delta Q
\end{equation}
\subsection{Extended thermodynamics} 
The phase space of thermodynamic values used in the first law (\ref{first_law}) and the very same relation (\ref{first_law}) can be generalized if one assumes that the cosmological constant $\L$ is not a fundamental one but a parameter which can be varied. The corresponding variation of gravitational Lagrangian over parameter $\L$ shows that the respective variation has a dimension of volume and might be called as a thermodynamic volume and it means that the parameter $\L$ might be identified with the thermodynamic pressure \cite{Kastor_CQG09}. The exact definitions of this pressure and conjugate thermodynamic volume are as follows:
\begin{equation}\label{press_vol}
P=-\frac{\Lambda}{8\pi}, \quad V=\left(\frac{\partial M}{\partial P}\right)_{S, Q^+_{\vp}}.
\end{equation}
Having used the definition of the thermodynamic volume (\ref{press_vol}) we can write the evident form for it:
\begin{eqnarray}
\nonumber V=\omega_{n-1}\left(\frac{r^n_{+}}{n}-\frac{\al+\L\e}{2\al}\left[\sum^{\frac{n-1}{2}}_{j=0}\frac{(-1)^jd^{2j}r^{n-2j}_{+}}{n-2j}+(-1)^{\frac{n+1}{2}}d^n\arctan\left(\frac{r_+}{d}\right)\right]+\right.\\\left.\frac{2^{p-2}(2p-1)^2\e q^{2p}}{\al(n-2p)}r^{n-\frac{2p(n-1)}{2p-1}}_{+}{_{2}F_{1}}\left(1,\frac{p(n-1)}{2p-1}-\frac{n}{2};\frac{p(n-1)}{2p-1}-\frac{n}{2}+1;-\frac{d^2}{r^2_{+}}\right)\right)
\end{eqnarray}
for odd $n$ and
\begin{eqnarray}
\nonumber V=\omega_{n-1}\left(\frac{r^n_{+}}{n}-\frac{\al+\L\e}{2\al}\left[\sum^{\frac{n}{2}-1}_{j=0}\frac{(-1)^jd^{2j}r^{n-2j}_{+}}{n-2j}+(-1)^{\frac{n}{2}}d^n\ln\left(\frac{r^2_+}{d^2}+1\right)\right]+\right.\\\left.\frac{2^{p-2}(2p-1)^2\e q^{2p}}{\al(n-2p)}r^{n-\frac{2p(n-1)}{2p-1}}_{+}{_{2}F_{1}}\left(1,\frac{p(n-1)}{2p-1}-\frac{n}{2};\frac{p(n-1)}{2p-1}-\frac{n}{2}+1;-\frac{d^2}{r^2_{+}}\right)\right)
\end{eqnarray}
for even $n$ correspondingly. We point out that the only substantial difference in the given relations is the $\sim\arctan(r_{+}/d)$ term for odd $n$ and the term $\sim\ln\left(r^2_{+}/d^2+1\right)$ for even $n$.

We also consider an additional intensive variable, which, as was shown \cite{Miao_EPJC16,Stetsko} is necessary when one tries to obtain Smarr relation. This additional variable and its conjugate value can be written as follows:
\begin{equation}\label{Psi_pot}
\Pi=\frac{\al}{8\pi\e}, \quad \Psi=\left(\frac{\partial M}{\partial\Pi}\right)_{S,Q^+_{\vp},P}
\end{equation}
Having used the latter relation we obtain the evident form for the value $\Psi$, namely for odd $n$ we arrive at:
\begin{eqnarray}\label{psi_odd}
\nonumber\Psi=\frac{\omega_{n-1}}{2}\left[\frac{(\al^2-\L^2\e^2)}{2\al^2}\left(\sum^{\frac{n-1}{2}}_{j=0}\frac{(-1)^jd^{2j}r^{n-2j}_{+}}{n-2j}+(-1)^{\frac{n+1}{2}}d^n\arctan\left(\frac{r_+}{d}\right)\right)-\frac{(\al+\L\e)^2}{4\al^2}\times\right.\\\left.\left(\sum^{\frac{n-1}{2}}_{j=1}(-1)^j\frac{2jd^{2j}}{n-2j}r^{n-2j}_{+}+(-1)^{\frac{n+1}{2}}d^n\left(n\arctan\left(\frac{r_+}{d}\right)-\frac{dr_+}{r^2_{+}+d^2}\right)\right)-\Psi_1\right],
\end{eqnarray}
whereas for even $n$ we derive:
\begin{eqnarray}\label{psi_even}
\nonumber\Psi=\frac{\omega_{n-1}}{2}\left[\frac{(\al^2-\L^2\e^2)}{2\al^2}\left(\sum^{\frac{n}{2}-1}_{j=0}\frac{(-1)^jd^{2j}r^{n-2j}_{+}}{n-2j}+(-1)^{\frac{n}{2}}\frac{d^n}{2}\ln\left(\frac{r^2_+}{d^2}+1\right)\right)-\frac{(\al+\L\e)^2}{4\al^2}\times\right.\\\left.\left(\sum^{\frac{n}{2}-1}_{j=1}(-1)^j\frac{2jd^{2j}}{n-2j}r^{n-2j}_{+}+(-1)^{\frac{n}{2}}\frac{d^n}{2}\left(n\ln\left(\frac{r^2_+}{d^2}+1\right)-\frac{2r^2_+}{r^2_{+}+d^2}\right)\right)-\Psi_1\right],\end{eqnarray}
where we have used the following notation in order to simplify the previous two relations:
 \begin{eqnarray}
\nonumber\Psi_1=\frac{2^{p-1}(2p-1)^2q^{2p}}{\al^2(2p-n)}r^{n-\frac{2p(n-1)}{2p-1}}_{+}\left[\L\e^2{_{2}F_{1}}\left(1,\frac{p(n-1)}{2p-1}-\frac{n}{2};\frac{p(n-1)}{2p-1}-\frac{n}{2}+1;-\frac{d^2}{r^2_+}\right)+\right.\\\left.\nonumber\frac{(2p-n)}{(n+2p-2)}\frac{(\al+\L\e)\e d^2}{r^2_{+}}{_{2}F_{1}}\left(2,\frac{p(n-1)}{2p-1}-\frac{n}{2}+1;\frac{p(n-1)}{2p-1}-\frac{n}{2}+2;-\frac{d^2}{r^2_+}\right)\right]+\\\nonumber\frac{2^{2p-3}(2p-1)^3\e^2 q^{4p}}{\al^2(4p-2pn-n)}r^{n-\frac{4p(n-1)}{2p-1}}_{+}\left[{_{2}F_{1}}\left(1,\frac{2p(n-1)}{2p-1}-\frac{n}{2};\frac{2p(n-1)}{2p-1}-\frac{n}{2}+1;-\frac{d^2}{r^2_+}\right)+\right.\\\left.\frac{(4p-2pn-n)d^2}{(2pn+n-2)r^2_+}{_{2}F_{1}}\left(2,\frac{2p(n-1)}{2p-1}-\frac{n}{2}+1;\frac{2p(n-1)}{2p-1}-\frac{n}{2}+2;-\frac{d^2}{r^2_+}\right)\right].
\end{eqnarray}
It should be pointed out that the main difference in the relations (\ref{psi_odd}) and (\ref{psi_even}) is of similar character as it was for the thermodynamic volume, related to the fact that odd and even dimensions have inverse tangential and logarithmic terms respectively.

Having extended thermodynamic phase space by additional variables such as $P$ and $\Pi$ we can write the extended first law in the form:
\begin{equation}\label{gen_first_law}
\delta M=T\delta S+\Phi^{+}_{\vp}\delta Q^{+}_{\vp}+\Phi_q\delta Q+V\delta P+\Psi\delta\Pi.
\end{equation}
Now, having utilized all the new thermodynamic variables and taking into account the equality (\ref{rel_entr_pot}) we obtain the Smarr relation, which can be represented as follows:
\begin{equation}\label{smarr_rel}
(n-2)M=(n-1)TS-2VP-2\Psi\Pi+\frac{1}{p}(3p-pn-1)\Phi_qQ,
\end{equation} 
and as a consequence for linear case ($p=1$) we have:
\begin{equation}
(n-2)M=(n-1)TS-2VP-2\Psi\Pi+(2-n)\Phi_qQ,
\end{equation}
which is typical for a black hole with linear Maxwell field. It should be emphasized that the Smarr relation (\ref{smarr_rel}) we have just obtained to great extent is grounded on the chosen form of scalar potential (\ref{sc_pot}) and this form seems to be preferable among the other possible variants.  
\subsection{Gibbs free energy and heat capacity}
Since we have introduced thermodynamic pressure and volume by the relations (\ref{press_vol}) the mass function $M$ would play the role of enthalpy \cite{Kastor_CQG09,Gunasekaran_JHEP12,Kubiznak_CQG2017}. It is known that more convenient function to analyze thermodynamic behaviour of a system, especially in case there is some critical behaviour, is the Gibbs free energy which can be introduced in the following way:
\begin{equation}
G=M-TS
\end{equation}
Utilizing the latter relation we can write the explicit relation for the Gibbs free energy, namely for odd $n$ we can write:
\begin{eqnarray}\label{Gibbs_odd}
\nonumber G=\frac{\omega_{n-1}}{16\pi}\left(\ve r^{n-2}_{+}+\frac{2\L}{n(n-1)}r^n_{+}+\frac{(\al+\L\e)^2}{2\al\e(n-1)}\left[\sum^{\frac{n-1}{2}}_{j=0}(-1)^j\frac{(2j-1)d^{2j}}{n-2j}r^{n-2j}_{+}+(-1)^{\frac{n+1}{2}}d^n\times\right.\right.\\\left.\left.\left((n-1)\arctan\left(\frac{r_+}{d}\right)-\frac{dr_+}{r^2_{+}+d^2}\right)\right]+\frac{2^p(2p-1)(4p-2pn-1)q^{2p}}{(n-1)(2p-n)}r^{n-\frac{2p(n-1)}{2p-1}}_{+}+G_1\right),
\end{eqnarray}
whereas for even $n$ we obtain:
\begin{eqnarray}\label{Gibbs_even}
\nonumber G=\frac{\omega_{n-1}}{16\pi}\left(\ve r^{n-2}_{+}+\frac{2\L}{n(n-1)}r^n_{+}+\frac{(\al+\L\e)^2}{2\al\e(n-1)}\left[\sum^{\frac{n}{2}-1}_{j=0}(-1)^j\frac{(2j-1)d^{2j}}{n-2j}r^{n-2j}_{+}+(-1)^{\frac{n}{2}}\frac{d^n}{2}\times\right.\right.\\\left.\left.\left((n-1)\ln\left(\frac{r^2_+}{d^2}+1\right)-\frac{2r^2_+}{r^2_{+}+d^2}\right)\right]+\frac{2^p(2p-1)(4p-2pn-1)q^{2p}}{(n-1)(2p-n)}r^{n-\frac{2p(n-1)}{2p-1}}_{+}+G_1\right),
\end{eqnarray}
and here we have used the following notation:
\begin{eqnarray}
\nonumber G_1=-\frac{2^{p-1}(2p-1)(4p-2pn-1)(\al+\L\e)q^{2p}}{(n-1)(2p-n)\al}r^{n-\frac{2p(n-1)}{2p-1}}_{+}{_{2}F_{1}}\left(1,\frac{p(n-1)}{2p-1}-\frac{n}{2};\frac{p(n-1)}{2p-1}-\frac{n}{2}+1;-\frac{d^2}{r^2_+}\right)\\\nonumber+\frac{2^{p}(2p-1)^2(\al+\L\e)d^2q^{2p}}{(n-1)(2p+n-2)\al}r^{n-2-\frac{2p(n-1)}{2p-1}}_{+}{_{2}F_{1}}\left(2,\frac{p(n-1)}{2p-1}-\frac{n}{2}+1;\frac{p(n-1)}{2p-1}-\frac{n}{2}+2;-\frac{d^2}{r^2_+}\right)\\\nonumber+\frac{2^{2p-3}(2p-1)^2(4pn-6p+1)\e q^{4p}}{(n-1)(4p-2pn-n)\al}r^{n-\frac{4p(n-1)}{2p-1}}_{+}{_{2}F_{1}}\left(1,\frac{2p(n-1)}{2p-1}-\frac{n}{2};\frac{2p(n-1)}{2p-1}-\frac{n}{2}+1;-\frac{d^2}{r^2_+}\right)\\+\frac{2^{2(p-1)}(2p-1)^3\e d^2q^{2p}}{(n-1)(2pn+n-2)\al}r^{n-2-\frac{4p(n-1)}{2p-1}}_{+}{_{2}F_{1}}\left(2,\frac{2p(n-1)}{2p-1}-\frac{n}{2}+1;\frac{2p(n-1)}{2p-1}-\frac{n}{2}+2;-\frac{d^2}{r^2_+}\right).
\end{eqnarray}
For the linear case ($p=1$) the Gibbs free energy can be written in a bit simpler from, in particular for odd $n$ we can write:
\begin{eqnarray}\label{gibbs_lin_odd}
\nonumber G=\frac{\omega_{n-1}}{16\pi}\left(\ve r^{n-2}_{+}+\frac{2\L}{n(n-1)}r^{n}_{+}+\frac{2(2n-3)}{(n-1)(n-2)}q^2r^{2-n}_{+}+\frac{(\al+\L\e)^2}{2\al\e(n-1)}\left[\sum^{\frac{n-1}{2}}_{j=0}(-1)^j\frac{(2j-1)d^{2j}}{n-2j}r^{n-2j}_{+}+\right.\right.\\\nonumber\left.\left.(-1)^{\frac{n+1}{2}}d^n\left((n-1)\arctan\left(\frac{r_+}{d}\right)-\frac{dr_{+}}{r^2_{+}+d^2}\right)\right]+\frac{(\al+\L\e)q^2}{\al(n-1)}\left[\sum^{\frac{n-5}{2}}_{j=0}(-1)^j\frac{(2n-2j-5)}{(2j+4-n)d^{2(j+1)}}r^{2j+4-n}_{+}+\right.\right.\\\left.\left.\nonumber\frac{(-1)^{\frac{n-3}{2}}}{d^{n-2}}\left((n-1)\arctan\left(\frac{r_+}{d}\right)-\frac{dr_{+}}{r^2_{+}+d^2}\right)\right]+\frac{\e q^4}{2\al(n-1)}\left[\sum^{\frac{3n-7}{2}}_{j=0}(-1)^j\frac{(4n-2j-7)}{(2j+6-3n)d^{2(j+1)}}r^{2j+6-3n}_{+}+\right.\right.\\\left.\left.\frac{(-1)^{\frac{3n-5}{2}}}{d^{3n-4}}\left((n-1)\arctan\left(\frac{r_+}{d}\right)-\frac{dr_{+}}{r^2_{+}+d^2}\right)\right]\right)
\end{eqnarray} 
and for even $n$ it takes the following form:
\begin{eqnarray}\label{gibbs_lin_even}
\nonumber G=\frac{\omega_{n-1}}{16\pi}\left(\ve r^{n-2}_{+}+\frac{2\L}{n(n-1)}r^{n}_{+}+\frac{2(2n-3)}{(n-1)(n-2)}q^2r^{2-n}_{+}+\frac{(\al+\L\e)^2}{2\al\e(n-1)}\left[\sum^{\frac{n}{2}-1}_{j=0}(-1)^j\frac{(2j-1)d^{2j}}{n-2j}r^{n-2j}_{+}+\right.\right.\\\nonumber\left.\left.(-1)^{\frac{n}{2}}d^n\left(\frac{(n-1)}{2}\ln\left(\frac{r^2_+}{d^2}+1\right)-\frac{r^2_{+}}{r^2_{+}+d^2}\right)\right]+\frac{(\al+\L\e)q^2}{\al(n-1)}\left[\sum^{\frac{n-6}{2}}_{j=0}(-1)^j\frac{(2n-2j-5)}{(2j+4-n)d^{2(j+1)}}r^{2j+4-n}_{+}+\right.\right.\\\left.\left.\nonumber\frac{(-1)^{\frac{n-2}{2}}}{d^{n-2}}\left(\frac{(n-1)}{2}\ln\left(\frac{d^2}{r^2_{+}}+1\right)+\frac{d^2}{r^2_{+}+d^2}\right)\right]+\frac{\e q^4}{2\al(n-1)}\left[\sum^{\frac{3n-8}{2}}_{j=0}(-1)^j\frac{(4n-2j-7)}{(2j+6-3n)d^{2(j+1)}}r^{2j+6-3n}_{+}+\right.\right.\\\left.\left.\frac{(-1)^{\frac{3n}{2}}}{d^{3n-4}}\left(\frac{(n-1)}{2}\ln\left(\frac{d^2}{r^2_+}+1\right)+\frac{d^2}{r^2_{+}+d^2}\right)\right]\right)
\end{eqnarray} 
It should be emphasized that the second sums in the relations (\ref{gibbs_lin_odd}) and (\ref{gibbs_lin_even}) (namely the sums $\sum^{\frac{n-5}{2}}_{j=0}$ and $\sum^{\frac{n-6}{2}}_{j=0}$) do not appear in the given above relations when $n=3$ and $n=4$ respectively.
\begin{figure}
\centerline{\includegraphics[scale=0.33,clip]{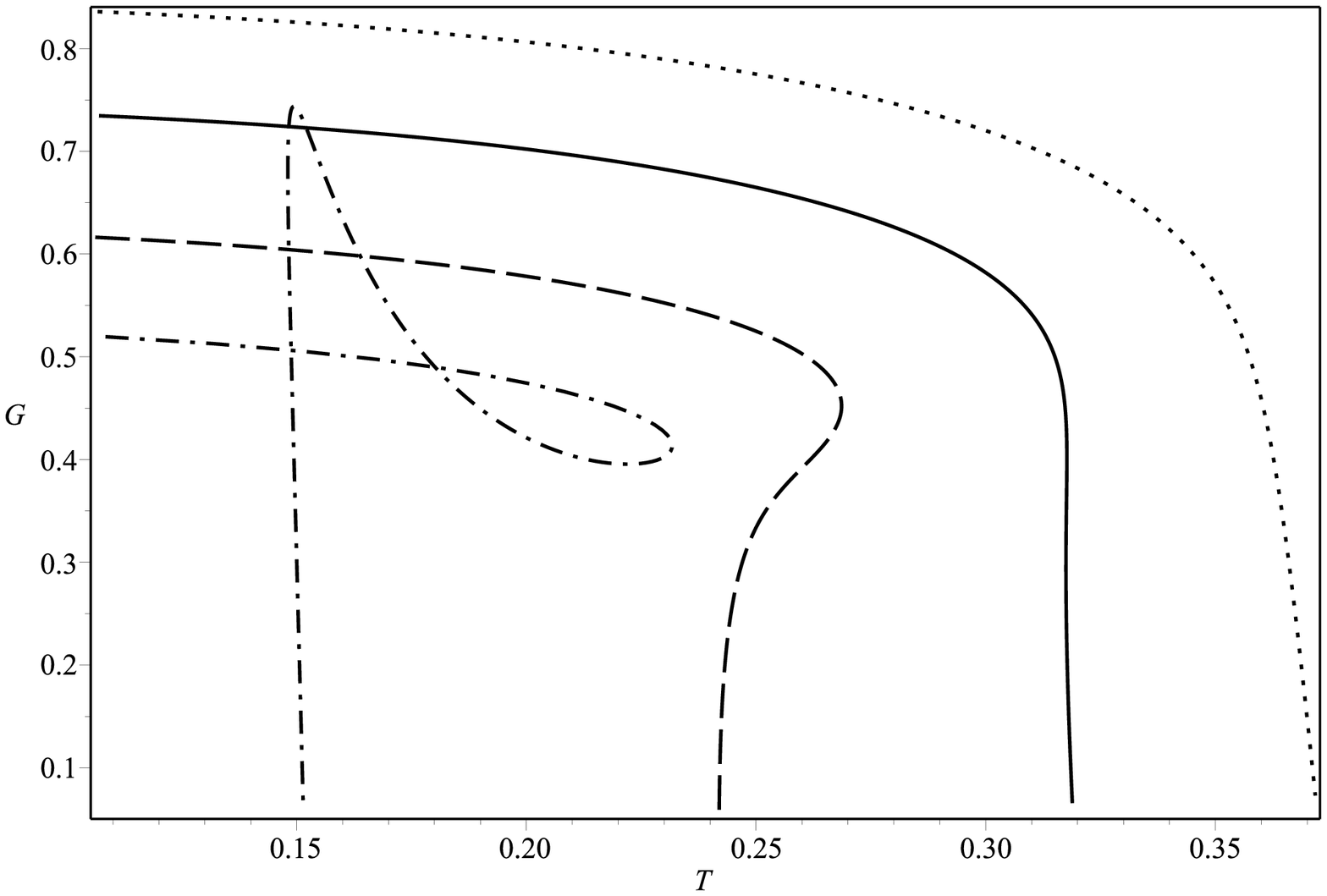}\includegraphics[scale=0.337,clip]{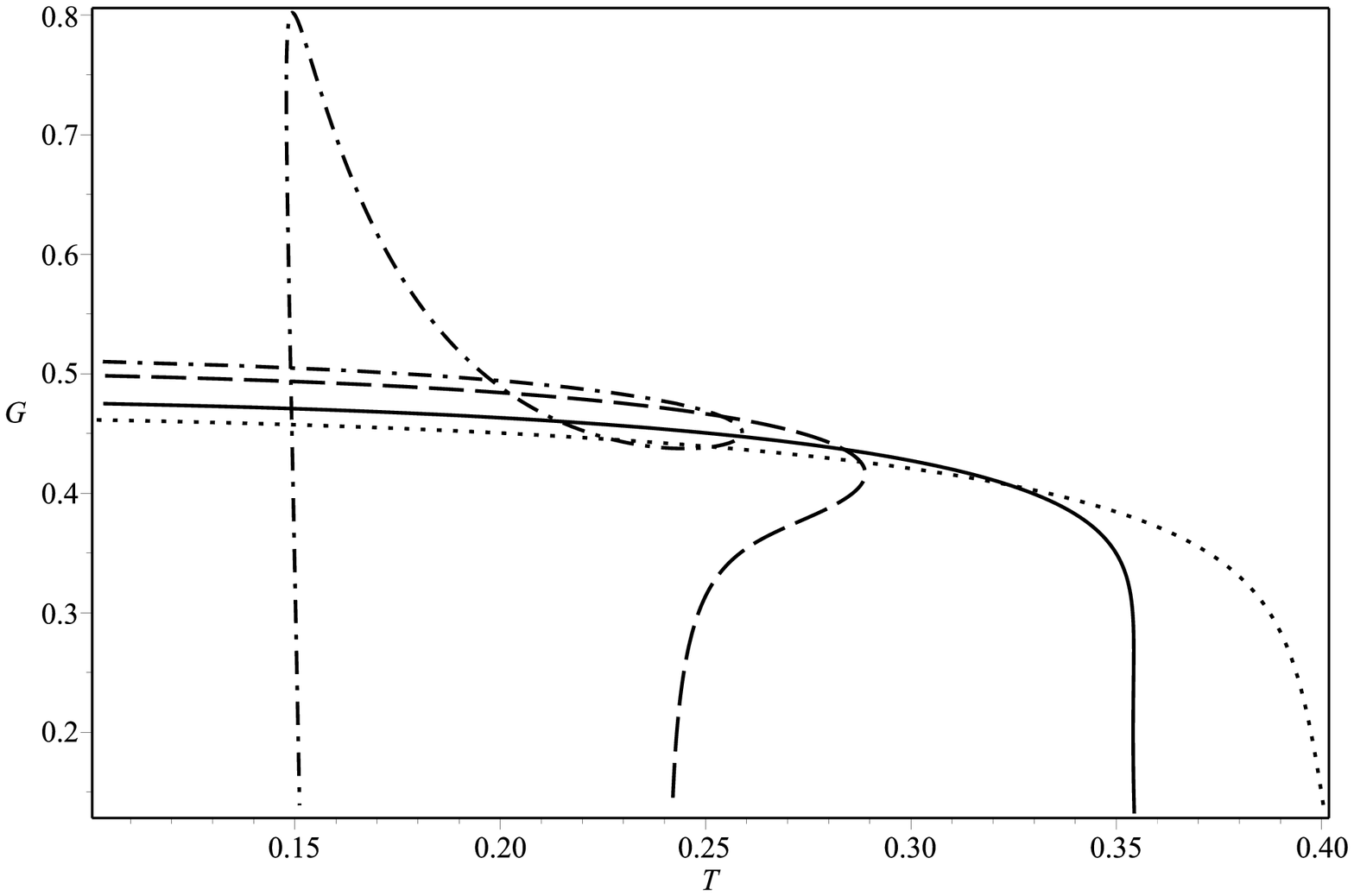}}
\caption{Gibbs free energy $G$ as a function of temperature $T$ for linear $p=1$ (the left graph) and conformal $p=\frac{n+1}{4}$ (the right graph) cases for several different values of pressure. For both graphs the solid curves represent almost critical value of pressure ($\L_c$), the dotted curves correspond to the pressure above the critical one, the dashed and dash-dotted curves correspond to the pressures below the critical values. For both graphs some parameters are the same, namely $n=4$, $\al=0.2$, $\e=0.4$, $\ve=1$ and $q=0.4$. For the left graph correspondence of curves is as follows (from the lowest the highest pressures): $\L=-0.8$, $\L=-3$, $\L_c=-5.7$ and $\L=-8$. For the right graph we have the following choice of the parameters: $\L=-0.8$, $\L=-3$, $\L_c=-7.4$ and $\L=-10$.}\label{Gibbs_1}
\end{figure}
 
We can also write the expressions for the Gibbs free energy in conformal case, namely, for odd $n$ we have:
\begin{eqnarray}\label{Gibbs_conf_odd}
\nonumber G=\frac{\omega_{n-1}}{16\pi}\left(\ve r^{n-2}_{+}+\frac{2\L}{n(n-1)}r^n_{+}+2^{\frac{n-3}{4}}n\frac{q^{\frac{n+1}{2}}}{r_+}+2^{\frac{n-7}{4}}\frac{(\al+\L\e)}{\al d}q^{\frac{n+1}{2}}\left((n-1)\arctan\left(\frac{r_+}{d}\right)-\frac{r_+d}{r^2_{+}+d^2}\right)\right.\\\nonumber\left.\frac{(\al+\L\e)^2}{2\al\e(n-1)}\left[\sum^{\frac{n-1}{2}}_{j=0}(-1)^j\frac{(2j-1)d^{2j}}{n-2j}r^{n-2j}_{+}+(-1)^{\frac{n+1}{2}}d^n\left((n-1)\arctan{\left(\frac{r_+}{d}\right)}-\frac{r_+d }{r^2_{+}+d^2}\right)\right]+\right.\\\left.2^{\frac{n-9}{2}}\frac{(n-1)\e}{\al}q^{n+1}\left[\sum^{\frac{n-1}{2}}_{j=0}(-1)^j\frac{(2(n-j)-1)}{2j-n}\frac{r^{2j-n}_{+}}{d^{2(j+1)}}+\frac{(-1)^{\frac{n+1}{2}}}{d^{n+2}}\left((n-1)\arctan\left(\frac{r_+}{d}\right)-\frac{r_+d}{r^2_{+}+d^2}\right)\right]\right).
\end{eqnarray}
It should be stressed that when $n=3$ the latter relation coincides with corresponding relation obtained from the relation for linear field (\ref{gibbs_lin_odd}) as it should be. For even $n$ the Gibbs free energy takes the form:
\begin{eqnarray}\label{Gibbs_conf_even}
\nonumber G=\frac{\omega_{n-1}}{16\pi}\left(\ve r^{n-2}_{+}+\frac{\L}{n(n-1)}r^n_{+}+2^{\frac{n-3}{4}}n\frac{q^{\frac{n+1}{2}}}{r_+}+2^{\frac{n-7}{4}}\frac{(\al+\L\e)}{\al d}q^{\frac{n+1}{2}}\left((n-1)\arctan\left(\frac{r_+}{d}\right)-\frac{r_+d}{r^2_{+}+d^2}\right)+\right.\\\nonumber\left.\frac{(\al+\L\e)^2}{2\al\e(n-1)}\left[\sum^{\frac{n}{2}-1}_{j=2}(-1)^j\frac{(2j-1)d^{2j}}{n-2j}r^{n-2j}_{+}+(-1)^{\frac{n}{2}}d^n\left(\frac{(n-1)}{2}\ln{\left(\frac{r^2_+}{d^2}+1\right)}-\frac{r^2_{+}}{r^2_{+}+d^2}\right)\right]+\right.\\\left.2^{\frac{n-9}{2}}\frac{(n-1)\e}{\al}q^{n+1}\left[\sum^{\frac{n}{2}-1}_{j=0}(-1)^j\frac{(2(n-j)-1)}{2j-n}\frac{r^{2j-n}_{+}}{d^{2(j+1)}}+\frac{(-1)^{\frac{n}{2}+1}}{d^{n+2}}\left(\frac{(n-1)}{2}\ln\left(\frac{d^2}{r^2_+}+1\right)+\frac{d^2}{r^2_{+}+d^2}\right)\right]\right).
\end{eqnarray}
The obtained above relations have some similarity with the linear case ($p=1$). The Fig.[\ref{Gibbs_1}]  represents the behaviour of the Gibbs free energy for linear $p=1$ and conformal  $p=\frac{n+1}{4}$ cases.  They show some similarity namely both of them have smooth behaviour for the pressures above the critical one ($P_c=-\L_c/8\pi$). The critical value of pressure $P_c$ (or cosmological constant $\L_c$) can be defined as the value when nonmonotonous  behvaviour in the dependence $T=T(r_+)$ disappears (see the Fig.[\ref{Temp_graph}]). For the pressures above this critical one the function $T=T(r_+)$ is monotonous, whereas for the pressures slightly below the critical values a specific maximum and minimum appear (Fig.[\ref{Temp_graph}]). It should be pointed out that in linear case ($p=1$) there  are some similarities with the behaviour of the function $G=G(T,P)$ for real gases and liquids, namely for the pressures above the critical value we have bigger values of the Gibbs free energy for the same temperatures, whereas for the conformal case the situation is opposite. Another important point we should emphasize is the fact that in both cases ($p=1$ and $p=(n+1)/4$) for the pressures substantially below the critical one a specific loop on the graphs of the function $G=G(T,P)$ appears, this situation in general is similar to the case of standard General Relativity or real condensed systems which are described by a Van der Waals-like equations of state. We also mention that for Van der Waals-systems (both gravitational and condensed matter) a first order phase transition appears for the pressures below the critical one and specific swallow-tail behaviour of the Gibbs free energy takes place. This swallow-tail loop appears immediately below the critical pressure and increase with corresponding decreasing of the pressure. We also point out that for some gravitational systems, namely for black holes with dilaton field additional phase transition of the zeroth order takes place. This zeroth order phase transition takes place when the pressure is below the critical one but the swallow-tail loop has not formed yet. In quite general features we have similar situation here, but to make a conclusion about type of phase transition deeper investigation of thermodynamic functions should be made and it will be performed elsewhere.

The other important function which brings considerable information about thermodynamic behaviour and especially about the phase transitions that might occur for a given system is the heat capacity:
\begin{equation}
C_{P}=T\left(\frac{\partial S}{\partial T}\right)_{P,\Pi,Q^{+}_{\vp},Q}=T\left(\frac{\partial S}{\partial r_+}\right)_{P}\left(\frac{\partial r_+}{\partial T}\right)_{P}
\end{equation}
It  should be pointed out that in the given above relation for the heat capacity we keep the parameters $Q^{+}_{\vp}$, $Q$ and $\Pi$ fixed and this fact allows us to write the evident form for the heat capacity as follows:
\begin{eqnarray}
\nonumber C_p=\frac{(n-1)\omega_{n-1}r^{n-2}_{+}}{4}\sqrt{U(r_+)W(r_+)}U'(r_+)\left(\frac{\ve(2-n)}{r^2_+}\left(1-\frac{(\al+\L\e)^2}{4\al^2}\right)+\frac{(\al-\L\e)^2}{2\al\e(n-1)}-\frac{2^p(4p-2pn-1)}{n-1}\times\right.\\\nonumber\left.r^{-\frac{2p(n-1)}{2p-1}}_{+}+\frac{(\al+\L\e)^2}{2\al\e(n-1)}\left[\sum^{\frac{n-1-\sigma}{2}}_{j=2}(-1)^j(1-2j)\frac{d^{2j}}{r^{2j}_+}+(-1)^{\frac{n+1-\sigma}{2}}\frac{d^{n+1-\sigma}((2+\sigma-n)d^2+(\sigma-n)r^{2}_{+})}{r^{n-1-\sigma}_{+}(r^2_{+}+d^2)^2}\right]+\right.\\\nonumber\left.\frac{2^{p-1}(\al+\L\e)}{(n-1)\al}q^{2p}r^{-\frac{2p(n-1)}{2p-1}}_{+}\left[(4p-2pn-1){_{2}F_{1}}\left(1,\frac{p(n-1)}{2p-1}-\frac{n}{2};\frac{p(n-1)}{2p-1}-\frac{n}{2}+1;-\frac{d^2}{r^2_{+}}\right)+\right.\right.\\\left.\left.\nonumber\frac{2(4p^2-1)(n-1)d^2}{(n+2p-2)r^2_{+}}{_{2}F_{1}}\left(2,\frac{p(n-1)}{2p-1}-\frac{n}{2}+1;\frac{p(n-1)}{2p-1}-\frac{n}{2}+2;-\frac{d^2}{r^2_{+}}\right)-\frac{8(2p-1)^2d^4}{(n+6p-4)r^{4}_{+}}\times\right.\right.\\\left.\left.\nonumber{_{2}F_{1}}\left(3,\frac{p(n-1)}{2p-1}-\frac{n}{2}+2;\frac{p(n-1)}{2p-1}-\frac{n}{2}+3;-\frac{d^2}{r^2_{+}}\right)\right]+\frac{2^{2p-3}(2p-1)^2\e }{\al(n-1)}q^{4p}r^{-\frac{4p(n-1)}{2p-1}}_{+}\left[\frac{(6p-4pn-1)}{2p-1}\times\right.\right.\\\left.\left.\nonumber{_{2}F_{1}}\left(1,\frac{2p(n-1)}{2p-1}-\frac{n}{2};\frac{2p(n-1)}{2p-1}-\frac{n}{2}+1;-\frac{d^2}{r^2_{+}}\right)+\frac{2(6p+1)(n-1)d^2}{(2pn+n-2)r^{2}_{+}}{_{2}F_{1}}\left(2,\frac{2p(n-1)}{2p-1}-\frac{n}{2}+1;\right.\right.\right.\\\left.\left.\left.\frac{2p(n-1)}{2p-1}-\frac{n}{2}+2;-\frac{d^2}{r^2_{+}}\right)-\frac{8(2p-1)d^4r^{-4}_+}{(2pn+n+4p-4)}{_{2}F_{1}}\left(3,\frac{2p(n-1)}{2p-1}-\frac{n}{2}+2;\frac{2p(n-1)}{2p-1}-\frac{n}{2}+3;-\frac{d^2}{r^2_{+}}\right)\right]\right)^{-1}.
\end{eqnarray}
We note that for simplification of the latter relation we have not substituted the explicit relations for the derivative $U'(r_+)$ and for the square root of the product of the metric functions  $\sqrt{U(r_+)W(r_+)}$. We also write the expression for the heat capacity in case of linear gauge field:
\begin{eqnarray}
\nonumber C_p=\frac{(n-1)\omega_{n-1}r^{n-2}_{+}}{4}\sqrt{U(r_+)W(r_+)}U'(r_+)\left(\frac{\ve(2-n)}{r^2_+}\left(1-\frac{(\al+\L\e)^2}{4\al^2}\right)+\frac{(\al-\L\e)^2}{2\al\e(n-1)}-\frac{2(3-2n)}{n-1}\times\right.\\\left.\nonumber q^2r^{2(1-n)}_{+}+\frac{(\al+\L\e)^2}{2\al\e(n-1)}\left[\sum^{\frac{n-1-\sigma}{2}}_{j=2}(-1)^j(1-2j)\frac{d^{2j}}{r^{2j}_{+}}+(-1)^{\frac{n+1-\sigma}{2}}\frac{d^{n+1-\sigma}}{r^{n-1-\sigma}_{+}}\frac{((2+\sigma-n)d^2+(\sigma-n)r^{2}_{+})}{(r^2_{+}+d^2)^2}\right]+\right.\\\nonumber\left.\frac{(\al+\L\e)q^2}{\al(n-1)}\left[\sum^{\frac{n-5-\sigma}{2}}_{j=0}(-1)^j\frac{(2j+5-2n)}{d^{2(j+1)}}r^{2(j+2-n)}_{+}+(-1)^{\frac{n-3-\sigma}{2}}\frac{d^{3+\sigma-n}}{r^{n-1+\sigma}_{+}}\frac{((2-\sigma-n)d^2-(n+\sigma)r^2_{+})}{(r^2_{+}+d^2)^2}\right]+\right.\\\left.\frac{\e q^4}{2\al(n-1)}\left[\sum^{\frac{3n-7-\sigma}{2}}_{j=0}(-1)^j\frac{(2j+7-4n)}{d^{2(j+1)}}r^{2(j+3-2n)}_{+}+(-1)^{\frac{3n-5-\sigma}{2}}\frac{d^{5-3n+\sigma}}{r^{n-1+\sigma}_{+}}\frac{((2-\sigma-n)d^2-(n+\sigma)r^2_{+})}{(r^2_{+}+d^2)^2}\right]\right)^{-1}.
\end{eqnarray}

\begin{figure}
\centerline{\includegraphics[scale=0.33,clip]{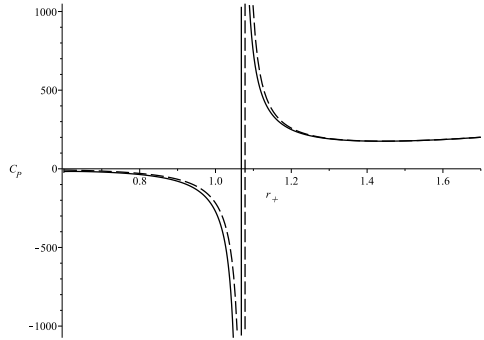}}
\caption{Comparison of heat capacities for linear (solid curve) and conformal (dashed curve) cases. All the fixed parameters are the same, namely $n=4$, $\ve=1$, $\al=0.2$, $\e=0.4$, $\L=-2$ and $q=0.2$. }\label{Heat_capac_fig}
\end{figure}

\begin{figure}
\centerline{\includegraphics[scale=0.33,clip]{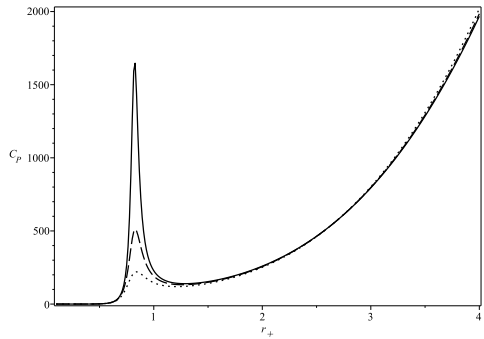}}
\caption{Heat capacity $C_P$ for conformal case when the discontinuity disappears. Increasing of the pressure $P$ lead to decreasing of the local maximum of heat capacity. The correspondence of the curves is as follows:  solid line corresponds to $\L=-2.1$, dashed curve corresponds to $\L=-2.15$ and the dotted curve corresponds to $\L=-2.25$. Other parameters are the same, namely $n=4$, $\ve=1$, $\al=0.2$, $\e=0.4$, and $q=0.4$. }\label{Heat_cap_2_fig}
\end{figure}

For the conformal case the heat capacity can be written in the form:
\begin{eqnarray}
\nonumber C_P=\frac{(n-1)\omega_{n-1}r^{n-2}_{+}}{4}\sqrt{U(r_+)W(r_+)}U'(r_+)\left(\frac{\ve(2-n)}{r^2_+}\left(1-\frac{(\al+\L\e)^2}{4\al^2}\right)+\frac{(\al-\L\e)^2}{2\al\e(n-1)}+\right.\\\nonumber\left.\frac{(\al+\L\e)^2}{2\al\e(n-1)}\left[\sum^{\frac{n-1-\sigma}{2}}_{j=2}(-1)^j(1-2j)\frac{d^{2j}}{r^{2j}_{+}}+\frac{(-1)^{\frac{n+1-\sigma}{2}}d^{n+1-\sigma}}{r^{n-1-\sigma}_{+}}\frac{((2+\sigma-n)d^2+(\sigma-n)r^{2}_{+})}{(r^2_{+}+d^2)^2}\right]+\right.\\\nonumber\left.\frac{2^{\frac{n-3}{4}}n q^{\frac{n+1}{2}}}{r^{n+1}_+}+\frac{2^{\frac{n-7}{4}}(\al+\L\e)q^{\frac{n+1}{2}}}{\al r^{n-1}_{+}(r^2_{+}+d^2)^2}((2-n)d^2-nr^2_{+})+\frac{2^{\frac{n-9}{2}}(n-1)\e q^{n+1}}{\al}\times\right.\\\left.\left[\sum^{\frac{n-1-\sigma}{2}}_{j=0}(-1)^j(2(j-n)+1)\frac{r^{2(j-n)}_+}{d^{2(j+1)}}+\frac{(-1)^{\frac{n+1-\sigma}{2}}}{d^{n+1-\sigma}}\frac{((2-\sigma-n)d^2-(\sigma+n)r^{2}_{+})}{r^{n-1+\sigma}_{+}(r^2_{+}+d^2)^2}\right]\right)^{-1}.
\end{eqnarray}
The comparison of the given above heat capacities is shown on the Fig.[\ref{Heat_capac_fig}]. The behaviour of the heat capacities is very similar, namely both of them have discontinuity which separate stable and unstable domains, for the horizon radii larger than these specific ones where discontinuity takes place the heat capacities are positive and it means that black holes are stable, whereas the black holes with radii smaller than these specific values are unstable because of the negative values of the heat capacities. It should be noted that for both cases $p=1$ and $p=(n+1)/4$ numerical values of heat capacities are very close in case all the fixed parameters take the same value. Since the function $T(r_+)$ might have two extrema points it means that the heat capacity possesses two points of discontinuity and one of those two points is depicted on the Fig.[\ref{Heat_capac_fig}], the second point of discontinuity takes place for the smaller values of the horizon radius $r_+$ and again it separates stable and unstable domains. We point out that the domain of instability is located in between those two points of discontinuity. When the pressure $P$ (or absolute value of $\L$) goes up the points of discontinuity become closer and for some specific value of pressure they merge and get transformed into a point of a local maximum of the heat capacity (Fig.[\ref{Heat_cap_2_fig}]). Further increasing of the pressure gives rise to the diminishing with the following disappearance  of the maximum, so for high pressures when the temperature of the black hole becomes monotonous function $T=T(r_+)$ the heat capacity gets transformed into a monotonous function also. 
\section{Conclusions}
In our work we have considered black hole solutions in a scalar-tensor theory with nonminimal derivative coupling between gravity and scalar field  and nonlinear electromagnetic field minimally coupled to gravity. We have taken the power law nonlinearity for the electromagnetic field Lagrangian in order to have possibility to obtain conformally invariant electrodynamics, but in our work we do not fix the parameter $p$ which characterizes degree of nonlinearity. As a result we have obtained solutions valid for different values of the parameter $p$, including $p=1$ which was previously derived in the paper \cite{Feng_PRD16} and $p=(n+1)/4$ as particular cases. Thus, the solutions we have obtained in our work are a generalization of the solutions derived in \cite{Feng_PRD16}. It should be pointed out that  the parameter $p$ is not completely arbitrary, to have reasonable behaviour of the electric field at the infinity and well-defined thermodynamic relations the parameter $p$  should satisfy the following condition $1/2<p<n/2$ and it is easy to see that linear $p=1$ and conformal $p=(n+1)/4$ cases fulfill the mentioned above condition.

 In this work we focus on static solutions and since we take into account cosmological constant $\L$ it gives us an opportunity to obtain topological black holes with different types of horizon surfaces, namely spherical ($\ve=1$), flat ($\ve=0$) and hyperbolic ($\ve=-1$) horizons. The obtained three types of solution have some common features, whereas the other ones might be distinct, namely at large distances ($r\rightarrow\infty$) all of them show anti-de Sitterian behaviour, at the same time for small distances ($r\rightarrow 0$) all three solutions represent different types of geometry. It should be pointed out that for the flat type geometry the obtained black hole's solution is the simplest one and is expressed in terms of elementary power-law functions of $r$ for any allowed value of $p$, whereas for other two types of geometry the general solution can be represented by hypergeometric functions which reduce to elementary ones only when the parameters of the hypergeometric functions are integer, in particular they are integer for linear $p=1$ and conformal cases $p=(n+1)/4$. Another important moment we should mention is the fact that in the regime of strong nonminimal coupling (when the term with factor $\e$ is relatively large in comparison with the term with the factor $\al$) the behaviour of metric functions for flat ($\ve=0$) and nonflat ($\ve=\pm 1$) geometries differs substantially. To investigate the question about  singularities of the obtained metrics we have considered the Kretschmann scalar (\ref{Kr_scalar}) which shows that there is just a coordinate singularity at the black hole's horizon $r_+$ and the only physical singularity is at the origin. It should be emphasized here that in the vicinity of the origin of coordinates the behaviour of the Kretschmann scalar is $\sim1/r^4$ for all types of solutions and arbitrary $n$ and all allowed values of $p$, and the leading term of this scalar (\ref{Kr_scal_origin}) does not depend on the integration constants $\mu$ and $q$ which are related to black hole's mass and charge respectively, this fact is specific peculiarity that takes place for the obtained black hole solutions and does not happen in the framework of standard General Relativity. At the infinity the Kretschmann scalar takes finite value which is completely defined by the parameters of coupling $\al$ and $\e$ and the dimension of space $n$, whereas the cosmological constant is not present here and this fact is the major difference from the situation in standard General Relativity where the presence of the constant $\L$ means that it appears in the Kretschmann scalar at the infinity.
 
 We have also obtained the potential of the gauge field and using Gauss law we have calculated black hole's charge. Here we point out that the electric potential, namely its change between the horizon and the infinity and the charge are extremely important for constructing of the first law of black hole's thermodynamics. It is worth being stressed that the mentioned above conditions imposed on the parameter $p$ are derived after careful analysis of the behaviour of the electric potential which should not diverge at the infinity. We also note that the same constraints on the parameter $p$ are obtained when we use Wald procedure to derive the first law of black hole's thermodynamics.
 
 Using standard geometrical relation we calculate black hole's temperature in general case and we also derive temperature for linear and conformal cases. The quite general trait of the temperature is the fact that it goes up almost linearly when the radius of horizon $r_+$ becomes large enough, similar behaviour of the temperature takes place in standard General Relativity when one takes into account the cosmological constant $\L$, but in our case instead of ``pure'' cosmological constant $\L$ an ``effective'' cosmological constant appears which depends on the ``pure'' value and the parameters of coupling $\al$ and $\e$. For small radii of horizon the temperature becomes negative. The Fig.[\ref{Temp_graph}] demonstrates the dependences $T=T(r_+)$ for some intermediate values of radii of horizon $r_+$, these graphs show that the temperature might be a nonmonotonous function of $r_+$ if the cosmological constant $\L$ (or the pressure we identify with) is lower than the specific value $\L_c$, whereas for the cosmological constant is larger than $\L_c$ the temperature becomes monotonic. The described behaviour of the temperature is similar to the situation that takes place for real systems in condensed matter physics, namely real gases and liquids. The nonmonotonic behaviour of temperature gives rise to the appearance of  phase transitions which separate stable and unstable phases.
 
 To obtain entropy of the black hole we have applied Wald's approach \cite{Wald_PRD93,Iyer_PRD94} which has shown its efficiency in scalar-tensor theories of gravity \cite{Feng_JHEP15, Feng_PRD16}. It should be pointed out that the relation for the entropy we have obtained differs from the well-known relation of standard General Relativity where the entropy is defined as a quarter of horizon area surface, namely we have obtained the relation (\ref{entropy}) which has this quarter of the area multiplied by some specific factor, caused by the nonminimal coupling and the obtained general relation for the entropy is the same for all types of the black holes we investigate. To derive the relation for entropy we have also introduced additional specific scalar ``charge'' $Q^+_{\vp}$ as it was proposed by Feng et al. \cite{Feng_PRD16} but this scalar ``charge'' can be chosen in different ways and similarly as it was done in our earlier work \cite{Stetsko} we take it in a different form than it was used in the original paper \cite{Feng_PRD16}. The evident form of this scalar ``charge'' we have taken allows us to derive Smarr relation (\ref{smarr_rel})  easily. Having introduced the entropy and the scalar charge $Q^{+}_{\vp}$ together with conjugate potential we have obtained  the first law of black hole thermodynamics. 
 
Having used the concept of thermodynamic pressure, which is related to the cosmological constant $\L$ \cite{Kastor_CQG09} we have developed the so called extended thermodynamics. Namely, we have derived the generalized first law (\ref{gen_first_law}) and the mentioned above Smarr relation (\ref{smarr_rel}). Having made Legendre transformation we have obtained the Gibbs free energy which is extremely useful for analysis of thermodynamic behaviour of the black hole. The Gibbs free energy has monotonous behaviour for the pressures above some specific value, which depends on the dimension of space, parameters of coupling $\al$ and $\e$, parameter of nonlinearity $p$, black hole's charge $q$ and type of geometry of the horizon. When the pressure decreases the Gibbs free energy becomes nonmonotonous and finally a closed loop appears that has some similarities with closed loop of a swallow-tail form which is typical for a Van der Waals systems or charged black holes in standard General Relativity \cite{Kubiznak_JHEP12,Gunasekaran_JHEP12}. The appearance of the loops gives rise to the conclusion about the existence of phase transitions and rich phase behaviour, but these issues need more detailed examination and will be considered elsewhere. Finally, we have calculated heat capacity for the black holes, namely we have written the evident form for it in general case as well as in particular cases linear $p=1$ and conformal $p=(n+1)/4$ electrodynamics. The obtained relations and as a consequence the corresponding graphs for them show some common features, namely for small pressures the heat capacity has two discontinuity points which separates stable and unstable phases, the increase of the pressure leads to merging of these discontinuities and its transformation into a peak which vanishes with corresponding increase of the pressure.  
 \section{Acknowledgements}
This work was partly supported by Project FF-30F (No. 0116U001539) from the Ministry of Education and Science of Ukraine.

\end{document}